\documentclass[12pt,a4paper]{article}
\usepackage{type1cm,amsmath,hangcaption,graphicx,indentfirst}
\usepackage[psamsfonts]{amssymb}
\numberwithin{equation}{section}

\begin{document}
\begin{titlepage}

 \renewcommand{\thefootnote}{\fnsymbol{footnote}}
\begin{flushright}
 \begin{tabular}{l}
 DESY 06-093\\
 hep-th/0606191\\
 \end{tabular}
\end{flushright}

 \vfill
 \begin{center}
 \font\titlerm=cmr10 scaled\magstep4
 \font\titlei=cmmi10 scaled\magstep4
 \font\titleis=cmmi7 scaled\magstep4
 \centerline{\titlerm Interactions for Winding Strings in Misner Space}
 \vskip 2.5 truecm

\noindent{ \large Yasuaki Hikida\footnote{E-mail:
yasuaki.hikida@desy.de}}
\bigskip

 \vskip .6 truecm
\centerline{\it DESY Theory Group, Notkestrasse 85, D-22603 Hamburg, Germany}

 \vskip .4 truecm

 \end{center}

 \vfill
\vskip 0.5 truecm

\begin{abstract}

We compute correlation functions of closed strings
in Misner space, a big crunch big bang universe.
We develop a general method for correlators 
with twist fields, which are relevant for the investigation 
on the condensation of winding tachyon.
We propose to compute the correlation functions by
performing an analytic continuation of the results
in ${\mathbb C}/{\mathbb Z}_N$ Euclidean orbifold.
In particular, we obtain a finite result for a 
general four point function of twist fields,
which might be important for the 
interpretation as the quartic term of the tachyon potential.
Three point functions are read off through the factorization,
which are consistent with the known results.

\end{abstract}
\vfill
\vskip 0.5 truecm

\setcounter{footnote}{0}
\renewcommand{\thefootnote}{\arabic{footnote}}
\end{titlepage}

\newpage

\section{Introduction}
\label{Intoduction}

Lorentzian (time-dependent) orbifolds attract much attention recently,
because they are solvable models describing strings in 
big crunch/big bang universes (see, for reviews, 
\cite{CCreview,Pioline4,Craps}).
Typical Lorentzian orbifolds, however, have
pathology associated with their big bang singularities.
It was shown in \cite{LMS1,LMS2,BCKR} that $2 \to 2$ scattering 
amplitude of untwisted states diverges due to graviton exchanges
near the big bang singularity.
Moreover, it was pointed out in \cite{HP} that a probe string 
induces a large backreaction due to the blue shift near the 
singularity, and produces a large black hole.
One way to resolve the singularity is
to add non-trivial directions such that the geometry has no region 
including the singular point \cite{FOS,LMS2,FM}.
This way of resolution, however, does not give any hint for 
stringy effects around the singularity.

The stringy way of resolution may come from perturbative or
non-perturbative effects. A perturbative way is the condensation
of winding strings, which could be massless \cite{Pioline2,Pioline3}
or tachyonic \cite{MS,NRS}.
It was suggested in \cite{Pioline2} that
the winding strings are pair-created due to the time-dependent 
background, and the backreaction may resolve the singularity.
On the other hand, it was conjectured in \cite{MS} that the big bang 
singularity is replaced by so-called tachyon state. 
The tachyon condensation was also investigated by D-instanton probe
in \cite{BKRS,She,HT}.
A non-perturbative way may be
given by Matrix model description of M-theory on time-dependent 
backgrounds. The backgrounds could be the ones with null-linear dilaton 
\cite{CSV} as well as Lorentzian orbifolds \cite{Robbins,MRS}.
In the dual description, the singular region corresponds to
the weak coupling region, where newly appearing non-Abelian fields 
seem to resolve the singularity. Recent developments along this 
direction are given, e.g., in 
\cite{Li,LS,DM1,Chen,IKO,LS2,CRS,CH,DM2,DMNT,LW,Chen2,IO,NP,KO,NPS}.

In this paper, we compute correlation functions of closed
strings in Misner space,%
\footnote{Some of the correlation functions were already 
computed in \cite{Pioline3}.
Though the correlators with less than two twist fields 
can be computed in operator formalism as in \cite{Pioline3},
it is difficult for those with more than two. 
It was proposed in \cite{Pioline3} that the three point functions
of twist fields can be obtained by utilizing the results of 
Nappi-Witten model in \cite{DAK,CFS,BDKZ}. 
However, it is obviously important to
construct a general formalism to compute the correlation functions in
the Lorentzian orbifold.} which is obtained as (1+1) dimensional
Minkowski space-time divided by a discrete boost.
The background includes a big crunch/big bang singularity,
and open and closed strings in the model have been
investigated in \cite{Nekrasov,Pioline1,Pioline2,Pioline3,HNP2,Tai}.
We develop a general formalism to compute the correlation functions,
in particular, we obtain the four point 
functions of winding strings, and find a finite result.
Considering the scattering of massless winding strings,
then the result suggests that the winding strings feel the 
singularity milder than the untwisted strings.
The four point function may also give sensible information on 
the quartic coupling of tachyonic fields due to its finiteness.

We propose to compute the general correlation functions 
in the following way.
First we define the Lorentzian orbifold, which describes
strings in Misner space, by an analytic continuation of
${\mathbb C}/{\mathbb Z}_N$ Euclidean orbifold theory
in the next section.
In particular, we will explain in detail how to map the 
spectra in Misner space and in
${\mathbb C}/{\mathbb Z}_N$ orbifold theory.
Once we have the map, then we only need to compute
general correlation functions in the
${\mathbb C}/{\mathbb Z}_N$ theory.
We start from several simple examples
to get a feeling of our method in section \ref{simple},
and then we move to general ones in section \ref{correlators}.
We mainly follow the technique developed in \cite{DFMS,HV},
where monodromy conditions around the twist fields are utilized
to compute correlators in the ${\mathbb C}/{\mathbb Z}_N$
orbifold. 
Four point functions are obtained in
subsection \ref{four} by utilizing the method,
and three point functions are deduced
from the four point functions in subsections \ref{ttt} and \ref{tut}.
In particular, the three point functions reproduce the results
in \cite{Pioline3}.
In section \ref{conclusion}, we conclude our results and
discuss the physical implications of the correlation functions.
In the appendix, we summarize
the formula of hyergeometric function.


\section{Misner space and ${\mathbb C}/{\mathbb Z}_N$ orbifold}
\label{map}

Misner space is defined as the (1+1) dimensional Minkowski space-time
with the identification of the discrete boost 
(${\mathbb R}^{1,1}/{\mathbb Z}$)
\begin{align}
 x^{\pm} &\sim e^{\pm 2 \pi \gamma} x^{\pm} ~,
 &x^{\pm} &= \frac{1}{\sqrt2} (x_1 \pm x_0) ~. 
 \label{Misner}
\end{align}
Since the boost does not mix the regions inside and outside
of the light-cone, the universe is divided by the lines $x^+ x^- =0$.
The regions with $x^+ x^- < 0$ are called as cosmological regions,
and the regions with $x^+ x^- > 0$ are called as whisker regions.
Changing the coordinates as
$x^{\pm} = \pm \frac{1}{\sqrt2} t e^{\pm \psi}$ for  $x^+ x^- < 0$ and 
$x^{\pm} = \frac{1}{\sqrt2} r e^{\pm \chi}$ for $x^+ x^- > 0$,
the metric is given by
\begin{align}
ds^2 &= -dt^2 + t^2 d \psi ^2 ~,
&ds^2 &=  d r^2 - r^2 d \chi ^2 ~,
\end{align}
where $\psi \sim \psi + 2 \pi \gamma$ and 
 $\chi \sim \psi + 2 \pi \gamma$. 
In the cosmological regions, the radius of space circle depends on
the time, and in particular, there is a big crunch/big bang
singularity at $t=0$. In the whisker regions, there are closed
time-like curves due to the periodicity of the time $\chi$.
These regions are conjectured to be excised from the universe 
by tachyon condensation \cite{Hagedorn}, which will 
be discussed later.
In the context of critical string theory, we implicitly add
$d$ extra flat directions, where $d=24$ for bosonic strings
and $d=8$ for superstrings.

In the following, we will connect this model with
a Euclidean orbifold (${\mathbb C}/{\mathbb Z}_N$)
with an enough large integer $N$.%
\footnote{In order to perform the path integral to compute
correlation functions, we may have to
Wick rotate the target space into a Euclidean space. In the orbifold
theory, the identification by a discrete boost has to be also mapped 
to the one by $2\pi/N$ rotation. It might be more transparent
to use an irrational $N$ for the purpose of the Wick rotation.
The assumption that $N$ is integer is only used in the next subsection
to make clear the normalization of untwisted states.}
The Euclidean orbifold is defined as 2 dimensional flat space
with the identification of $2\pi / N$ rotation
\begin{align}
 x^{\pm} &\sim e^{\pm 2 \pi i / N} x^{\pm} ~,
 &x^{\pm} &= \frac{1}{\sqrt2} (x_1 \pm i x_2) ~.
 \label{Orbifold}
\end{align}
The condition for Misner space \eqref{Misner} may be
obtained just by replacing $1/N \leftrightarrow - i \gamma$
and $ x_2 \leftrightarrow - i x_0$.
In the next section, we extend the orbifold procedure
in the untwisted sector from the Euclidean orbifold case into 
the Lorentzian orbifold case. 
We should take a care on the fact that
infinitely many images of the discrete boost should be summed over 
in the Lorentzian orbifold.
In subsection \ref{twist}, we propose how to define the
map between the spectra of both the orbifolds.
In fact, the twisted sector is found to be more subtle,
and we need a trick to perform the analytic continuation.

\subsection{The untwisted sector}
\label{untwist}

We start from the states in the untwisted sector, which 
correspond to tachyon or graviton fields propagating the 
universe. Thus, the correlators involving these states
provide information about the background geometry.
In the 2 dimensional flat space, the tachyon field may be defined as
\begin{align}
V_p &= e^{i \bar p X^+ + i p X^- } ~,
 &\langle V_p (\infty) V_{p '} (0) \rangle 
  &= (2 \pi)^2 \delta^{(2)}(p - p ') ~,
  \label{tachyon}
\end{align} 
or in the bra and ket form as
\begin{align}
 \langle p , \bar p | p ' , \bar p ' \rangle
  = (2 \pi)^2 \delta^{(2)}(p - p ') ~.
\end{align}
Here we defined $p = \frac{1}{\sqrt2}(p_1 + i p_2)$
and $\bar p = \frac{1}{\sqrt2} (p_1 - i p_2)$.
In the ${\mathbb C}/{\mathbb Z}_N$ orbifold theory,
the invariant state is given by summing over the
orbifold images as
\begin{align}
 | p ,\bar p \rangle_N &= 
  \frac{1}{N} \sum_{n=0}^{N-1} 
  | e^{\frac{2 \pi i n}{N}} p , e^{\frac{- 2 \pi i n}{N}} \bar p \rangle ~,
&{}_N\langle p , \bar p | p ' , \bar p ' \rangle_N
  &= (2 \pi)^2 \frac{1}{N}
 \sum_{n=0}^{N-1} \delta^{(2)}(p - e^{\frac{2 \pi i n}{N}} p ') ~.
\end{align}
We can similarly define the orbifold invariant states in the Misner
space as
\begin{align}
\begin{aligned}
 | p ,\bar p \rangle_{\gamma} &= 
  \frac{1}{Z} \sum_{n \in {\mathbb Z}} 
  | e^{ 2 \pi n \gamma} p^+ , e^{ - 2 \pi n \gamma} p^- \rangle ~,
\\
{}_{\gamma} \langle p , \bar p | p ' , \bar p ' \rangle_{\gamma}
  &= (2 \pi)^2 \frac{1}{Z}
 \sum_{n \in {\mathbb Z}} 
  \delta (p^+ - e^{ 2 \pi n \gamma} {p^+} ') 
  \delta (p^- - e^{ - 2 \pi n \gamma} {p^-} ') ~,
\label{untwisted}
\end{aligned}
\end{align}
where $p^{\pm} = \frac{1}{\sqrt2}(p_1 \pm p_0)$.
Note that we divide by the infinite factor $Z=\sum_{n}1$.
The completeness conditions are
\begin{equation}
1 = \int \frac{dp d \bar p}{(2 \pi)^2} | p ,\bar p \rangle \langle p ,\bar p |
  = \int \frac{dp d \bar p}{(2 \pi)^2} | p ,\bar p \rangle_N {}_N\langle p ,\bar p |
  = \int \frac{dp^+ dp^-}{(2 \pi)^2} | p^+ ,p^- \rangle_{\gamma} {}_{\gamma}\langle p^+ , p^- | ~,
\label{complete}
\end{equation}
which will be used to factorize four point functions
into the product of two three point functions. 
Keeping these relations in mind, we will adopt the integral over 
the complex plane of $p$ and the basis for the flat space.

The standard basis in Misner space \cite{Nekrasov,BCKR} is 
not the one in \eqref{untwisted}, but the both can be mapped
into each other as follows. Let us define as
\begin{align}
 | q ,n \rangle_N = \frac{1}{N} \int_0^{N} ds
   | e^{\frac{2 \pi i s}{N}} p , e^{\frac{- 2 \pi i s}{N}} \bar p \rangle
    e^{- 2 \pi i n s} ~,
\end{align}
with $|p|=q$, which is also invariant under the orbifold action.
The measure is given by
$\sum_{n \in {\mathbb Z}} \int \frac{q dq}{(2 \pi)}$.
The two point function can be computed as
\begin{align}
 {}_N \langle q , n | q ' , n ' \rangle_N
  = \frac{1}{N^2} \int_0^{N} ds \int_0^{N} ds'
   (2 \pi)^2 \delta^{(2)} ( e^{\frac{2 \pi i s}{N}} q - e^{\frac{2 \pi i s '}{N}} q ' )
   e^{2 \pi i (n s - n' s')} ~.
\end{align}
The phase factors are absorbed by the constant
shift of $s$ and $s'$ implicitly.
Changing the variables as $\theta^{\pm} = s \pm s'$, we obtain
\begin{align}
\frac{1}{2 N^2} \int_0^{ 2 N} d \theta^+ \int_{0}^{N} d \theta^-
   (2 \pi)^2 \frac{1}{q} \delta ( q - q') 
 \delta ( {\textstyle \frac{2 \pi}{N}} \theta^- )
   e^{\pi i ( (n - n') \theta^+ + (n + n') \theta^- )} ~.
\end{align}
The integral over $\theta^+$ gives the delta function
$2 N \delta_{n,n'}$, and the integral over $\theta^-$
picks up the delta function and leads $N/(2\pi)$.
Therefore, we obtain the inner product of this basis as
\begin{align}
 {}_N \langle q , n | q ' , n ' \rangle_N
  = (2 \pi) \frac{1}{q} \delta (q - q') \delta_{n,n'} ~.
  \label{inner}
\end{align}
The completeness condition can be also shown as
\begin{align}
\begin{aligned}
 &\sum_n \int_{0}^{\infty} \frac{q d q}{2 \pi}
  | q  , n  \rangle_N {}_N \langle q , n | \\
  &= \frac{1}{N^2}
    \int_{0}^{\infty} \frac{q d q}{2 \pi} \int_0^N ds \int_0^N ds'
    | q e^{\frac{2 \pi i s}{N}} , q e^{- \frac{2 \pi i s}{N}} \rangle
    \langle q e^{\frac{2 \pi i s '}{N}} , q e^{- \frac{2 \pi i s '}{N}} | 
     \sum_m \delta (s - s'+m )\\
  &= \frac{1}{N}\int_{0}^{\infty} \frac{q d q}{(2 \pi)^2} \int_0^{2 \pi} d \psi 
   \sum_m | q e^{ i \psi} , q e^{- i \psi} \rangle
    \langle q e^{ i \psi + \frac{2 \pi i m}{N}} ,
            q e^{ - i \psi - \frac{2 \pi i m}{N}} | = 1 ~.
\end{aligned}
\label{complete2}
\end{align}
In the first equality, the sum over $n$ is taken, which leads to the delta function.
In the second equality, the integral over $s'$ picks up the delta
function, and the variable changes as $\psi = 2 \pi s / N$.
The last equality is the consequence of \eqref{complete}.
With a simple generalization, we can define the basis in
Misner space as
\begin{align}
 | q ,n \rangle_{\gamma} = \frac{1}{Z '} \int_{- \infty}^{\infty} ds
   | e^{ 2 \pi s \gamma} p^+ , e^{ - 2 \pi s \gamma} p^- \rangle
    e^{- 2 \pi i n s} ~,
\end{align}
which is the same as the one in \cite{Nekrasov,BCKR}. 
The inner product and the completeness
condition are satisfied with
$Z ' = \sqrt{\sum_n 1 / \gamma}$.
In this way, we have shown that 
the result does not depend on the choice of the basis.

\subsection{The twisted sectors}
\label{twist}

As mentioned above, it is a subtle problem to define the Hilbert 
space of twisted sectors in Misner space.
In order to clarify the problem, we first study
the well-known spectrum in the
${\mathbb C}/{\mathbb Z}_N$ orbifold theory.
In the twisted sector, the closed strings have to
satisfy the twisted boundary conditions
\begin{align}
 X^{\pm} (e^{2 \pi i} z , e^{- 2 \pi i} \bar z )
 = e^{\pm \frac{2 \pi i k}{N} } X^{\pm} (z , \bar z)
 \label{tbc}
\end{align}
with an integer $k \neq 0$.%
\footnote{We should 
assume $|k| < N$ in the ${\mathbb C}/{\mathbb Z}_N$ theory, 
but we can extend to any integer number in Misner space case.}
Therefore, the mode expansion in the $k$-th twisted sector
is given by\footnote{We set $\alpha ' =2$ throughout this paper.}
\begin{align}
 X^{\pm} (z,\bar z) = i \sum_n \frac{1}{(n \mp \frac{k}{N})}
                \frac{\alpha^{\pm}_n}{z^{n \mp \frac{k}{N}}}
             + i \sum_n \frac{1}{(n \pm \frac{k}{N})}
                \frac{\tilde \alpha^{\pm}_n}{{\bar z}^{n \pm \frac{k}{N}}} ~,
\end{align}
where the mode operators satisfy
\begin{align}
 [ \alpha^-_m , \alpha^+_n ] &= (m + \frac{k}{N}) \delta_{m+n} ~,
 &[ \tilde \alpha^-_m , \tilde \alpha^+_n ] &= (m - \frac{k}{N}) \delta_{m+n} ~.
\label{comm_rel}
\end{align}
The Hilbert space is generated by acting the negative mode operators
to the states with quasi-zero modes (we assume $k > 0$ from now on)
\begin{align}
 | n , \tilde n , k \rangle_N =
  (\alpha^{+}_{0})^n  (\tilde \alpha^{-}_{0})^{\tilde n} | k \rangle_N ~,
\label{vac_twist_ket}
\end{align}
where the vacuum state has the property as
\begin{align}
  \alpha^+_{n > 0} |k \rangle_N &= \alpha^-_{n \geq 0} | k \rangle_N = 0 ~,
  &\tilde \alpha^+_{n \geq 0} | k \rangle_N 
     &= \tilde \alpha^-_{n > 0} | k \rangle_N = 0 ~.
     \label{vac_twist}
\end{align}
Note that the (quasi-)zero mode part is different from one in the untwisted 
sector since the quasi-zero modes satisfy non-trivial commutation 
relations \eqref{comm_rel}.

Naively we expect that the spectrum of twisted sectors in Misner space
is given just by
replacing $1/N \leftrightarrow - i \gamma$, $X_2 \leftrightarrow -i X_0$.
Indeed, we can obtain in this way the boundary conditions
\begin{align}
 X^{\pm} (e^{2 \pi i} z , e^{- 2 \pi i} \bar z )
 = e^{\pm 2 \pi k \gamma} X^{\pm} (z , \bar z) ~,
\end{align}
and the mode expansions
\begin{align}
 X^{\pm} (z,\bar z) = i \sum_n \frac{1}{(n \pm i k \gamma)}
                \frac{\alpha^{\pm}_n}{z^{n \pm i k \gamma}}
             + i \sum_n \frac{1}{(n \mp i k \gamma)}
                \frac{\tilde \alpha^{\pm}_n}{{\bar z}^{n \mp i k \gamma}} 
\end{align}
with
\begin{align}
 [ \alpha^-_m , \alpha^+_n ] &= (m - i k \gamma) \delta_{m+n} ~,
 &[ \tilde \alpha^-_m , \tilde \alpha^+_n ] &= (m + i k \gamma) \delta_{m+n} ~.
\label{comm_rel2}
\end{align}
However, the definition of the Hilbert space is a subtle problem.
As was pointed out in \cite{Nekrasov}, if we define the Hilbert space
as in the ${\mathbb C}/{\mathbb Z}_N$ orbifold, 
then there is no physical state. This is because
the Virasoro generators are implicitly defined as
\begin{align}
\begin{aligned}
 L_0 &=  - \frac{1}{2} i k \gamma (1 + i k \gamma ) + \alpha_0^+ \alpha_0^-
 + \sum_{n > 0} (\alpha^+_{-n} \alpha^-_n + \alpha^-_{-n} \alpha^+_n ) ~,\\
 \tilde L_0 &= - \frac{1}{2} i k \gamma (1 + i k \gamma) 
  + \tilde \alpha_0^- \tilde \alpha_0^+
 + \sum_{n > 0} (\tilde \alpha^+_{-n} \tilde \alpha^-_n 
  + \tilde \alpha^-_{-n} \tilde \alpha^+_n ) ~,
\end{aligned}
\end{align}
and they are imaginary for every states in the Hilbert space.

A solution was proposed in \cite{Pioline1}.
In terms of Virasoro generators, the change should be made as
\begin{align}
\begin{aligned}
 L_0 &= - \frac{1}{2} (i k \gamma )^2  + 
 \frac{1}{2}( \alpha_0^+ \alpha_0^- +  \alpha_0^- \alpha_0^+ )
 + \sum_{n > 0} (\alpha^+_{-n} \alpha^-_n + \alpha^-_{-n} \alpha^+_n ) ~,\\
 \tilde L_0 &= - \frac{1}{2} ( i k \gamma )^2
  +  \frac{1}{2}( \tilde \alpha_0^- \tilde \alpha_0^+
                + \tilde \alpha_0^+ \tilde \alpha_0^- )
 + \sum_{n > 0} (\tilde \alpha^+_{-n} \tilde \alpha^-_n 
  + \tilde \alpha^-_{-n} \tilde \alpha^+_n ) ~,
\end{aligned}
\end{align}
which means that the vacuum state is not the one in
\eqref{vac_twist} but the eigenstates as
\begin{align}
\begin{aligned}
 \frac{1}{2}( \alpha_0^+ \alpha_0^- +  \alpha_0^- \alpha_0^+ )
 | \omega^2 , {\tilde \omega}^2 ,  k \rangle_{\gamma}
 = \omega^2 | \omega^2 , {\bar \omega}^2  , k \rangle_{\gamma} ~, \\
 \frac{1}{2}( \tilde \alpha_0^+ \tilde \alpha_0^- +  \tilde \alpha_0^- \tilde \alpha_0^+ )
 | \omega^2 , {\tilde \omega}^2 , k \rangle_{\gamma}
 = {\tilde \omega}^2 | \omega^2 , {\tilde \omega}^2  , k \rangle_{\gamma} ~.
\end{aligned}
\end{align}
By means of the vacuum state \eqref{vac_twist}, this state may be
expressed as
\begin{align}
 | \omega^2 , {\tilde \omega}^2 ,  k \rangle_{\gamma}
 =   (\epsilon_+ \alpha^{+}_{0})^{ - \frac{\omega^2}{i k \gamma} - \frac12} 
    (\tilde \epsilon_- \tilde \alpha^{-}_{0})^{- \frac{{\tilde \omega}^2}{i k \gamma} - \frac12} 
   | k \rangle_{\gamma} ~,
\label{vac_Misner}
\end{align}
which has the extra labels 
$\epsilon_+ , \tilde \epsilon_- = \pm 1$ as explained below.

Restricting to the quasi-zero mode sector, we can see the
physical picture of these states. For the massive case
$\omega^2 = \tilde \omega^2 = - {\cal M}^2 < 0$, the solutions with 
$\alpha^{\pm}_0 = \pm \epsilon {\cal M}$ and
$\tilde \alpha^{\pm}_0 = \pm \tilde \epsilon \tilde {\cal M}$
lead to (now we set $z = e^{i(\tau + \sigma)}$
and $\bar z = e^{i(\tau - \sigma)}$)
\begin{align}
X^{\pm} (\tau,\sigma) &=
 \pm \frac{ 2 {\cal M}}{ k \gamma} e^{\pm k \gamma \sigma} \sinh k \gamma \tau
 ~,
&X^{\pm} (\tau,\sigma ) &=
\frac{ 2 {\cal M}}{k \gamma} e^{\pm k \gamma \sigma} \cosh k \gamma \tau ~.
\end{align}
The first solution is for $\epsilon = \tilde \epsilon = + 1$,
and it describes a string winding around $\psi$-cycle and wrapping
the whole cosmological regions. This type of winding string
may be used to study the singularity by the winding string 
probe. The second solution is for
$\epsilon = - \tilde \epsilon = + 1$, and it describes a
string winding around the periodic time $\chi$ and the string
exists from a point $r_0 = 2 \sqrt2 {\cal M}/( k \gamma )$
to a spatial infinity.

For the tachyonic case
$\omega^2 = \tilde \omega^2 > 0$, the solutions with 
$\alpha^{\pm}_0 = \epsilon \omega$ and
$\tilde \alpha^{\pm}_0 = \tilde \epsilon \tilde \omega$
lead to
\begin{align}
X^{\pm} (\tau,\sigma) &=
 \frac{ 2 \omega}{ k \gamma} e^{\pm k \gamma \sigma} \sinh k \gamma \tau
 ~,
&X^{\pm} (\tau,\sigma ) &=
 \pm \frac{ 2 \omega}{k \gamma} e^{\pm k \gamma \sigma} \cosh k \gamma \tau ~.
\label{winding_tachyon}
\end{align}
The first solution is for $\epsilon = \tilde \epsilon = + 1$,
and it describes a string wrapping the whole whisker regions. 
In the context of tachyon condensation, this type of tachyon 
is interesting since the condensation may remove
the whisker regions \cite{Hagedorn}, and at the same time 
resolve the big bang singularity in the cosmological 
region \cite{HT}.
The second solution is for
$\epsilon = - \tilde \epsilon = + 1$, and it describes
strings pair created at a specific time 
$t_0 = 2 \sqrt2 \omega/( k \gamma )$.

We need a special care on dealing with the state \eqref{vac_Misner},
since it is given by the complex power of the operators. 
In order to define this complex power, we utilize the Mellin transform
as in \cite{Pioline3}
\begin{align}
 (\epsilon_+ \alpha^{+}_{0})^{- \frac{\omega^2}{i k \gamma} - \frac12}
  &= \frac{1}{\Gamma(\frac{\omega^2}{i k \gamma} + \frac12)}
  \int_0^{\infty} d x x^{\frac{\omega^2}{i k \gamma} - \frac12} 
   e^{- \epsilon_+ \alpha^+_0 x} ~,
  &e^{- \alpha^+_0 y} &= \sum_{n=0}^{\infty} 
   \frac{(\alpha^+_0)^n (- y)^n}{n!} ~.
   \label{Mellin}
\end{align}
In other words, we define the exponential form of the state as
\begin{align}
 | y , \tilde y , k \rangle_N =  \sum_{n=0}^{\infty} 
   \frac{(- y)^n ( - \tilde y)^{\tilde n}}{n! \tilde n !} |n,\tilde n , k \rangle_N ~,
\label{xyk}
\end{align}
and perform the Wick rotation, namely, replace
$X_2 \leftrightarrow -i X_0$ and
$1/N \leftrightarrow - i \gamma$.
Moreover, we may perform the Mellin transformation \eqref{Mellin},
since the above basis \eqref{xyk} is not the eigenstate of $L_0$ and $\tilde L_0$.
In summary, it is enough to compute the correlators of the states 
\eqref{xyk} in the ${\mathbb C}/{\mathbb Z}_N$ orbifold theory
since the correlators in Misner space can be obtained through
the map.


\section{Simple examples}
\label{simple}

Before moving to the full analysis of correlation functions,
we start from some simple calculations.
In the next subsection, we compute correlation functions
with two twist fields.
Since we can map two twisted fields to in and out states,
we can apply operator formalism to this type of correlators.
In order to compute correlators with more than two twist
fields, we have to apply more general formalism, which will 
be developed in the next section. In subsection \ref{classical},
we compute classical three point functions of twisted fields, 
which are given by the overlaps of corresponding wave functions.
The both correlators were already computed in \cite{Pioline3},
but we rederive them in a bit different way
such that our prescription is manifest.

\subsection{Correlation functions in operator formalism}
\label{operator}

First we compute two point functions in the twisted sector,
even though the two pint functions are nothing but the normalization
of the states.
For the out states, we define the bra states as
\begin{align}
 {}_N \langle n , \tilde n , k | =
  {}_N \langle  k | (\alpha^{-}_{0})^n  (\tilde \alpha^{+}_{0})^{\tilde n} ~,
  \label{bra_state}
\end{align}
where the vacuum state satisfies
\begin{align}
  {}_N \langle k | \alpha^+_{n \leq 0}  
  &= {}_N \langle k | \alpha^-_{n < 0}  = 0 ~,
  & {}_N \langle k | \tilde \alpha^+_{n < 0}  
     &= {}_N \langle k |\tilde \alpha^-_{n \leq 0} = 0
     \label{vac_twist_bra}
\end{align}
and
\begin{align}
{}_N \langle  k |  k' \rangle _{N} = \delta_{k,k'} ~. 
\end{align}
With the help of the commutation relations for the operators \eqref{comm_rel}, 
we find
\begin{align}
 {}_N \langle n , \tilde n , k | n , \tilde n , k \rangle_{N}
  &= n! \tilde n ! \left( \frac{k}{N} \right)^{n+ \tilde n} ~,
   &{}_N \langle y , \tilde y , k | x ' , \tilde y ' , k \rangle_{N} 
  &= e^{ \frac{k}{N} (y y ' + \tilde y \tilde y' ) } ~.
  \label{twistmetric}
\end{align}
In the Misner space case, $1/N$ is replaced by $- i \gamma$.
Furthermore, we have to perform the Mellin transformation \eqref{Mellin},
in order to change the basis into the one in \eqref{vac_Misner}.
In this case, we can explicitly integrate over $y,y',\tilde y,\tilde y'$, 
which reads
\begin{align}
  {}_{\gamma} \langle \omega^2 , \tilde \omega^2 , k | 
                      \omega '{}^2, {\tilde \omega} '{}^2 , k \rangle_{\gamma} 
 = \frac{\xi (-i)^2 (i k \gamma)^{-\frac{\omega^2}{i k \gamma}
          -\frac{\tilde \omega^2}{i k \gamma} - 1}}
            {\Gamma (\frac{\omega^2}{i k \gamma}+\frac12)
             \Gamma (\frac{\tilde \omega^2}{i k \gamma}+\frac12)}   
          \delta ( \omega^2 - {\omega '}^2 ) \delta (\tilde \omega^2 - {\tilde \omega}'{}^{2})
\end{align}
with $\xi =(\epsilon_+ \epsilon_-)^{-\frac{\omega^2}{i k \gamma} - \frac12 }
  (\tilde \epsilon_- \tilde \epsilon_+)^{ 
-\frac{\tilde \omega^2}{i k \gamma} - \frac12 }$.
It might be convenient to change the normalization of the 
two point function such that the coefficient of the delta function
is one.

In order to compute more hight point functions, we need to insert 
vertex operators. In the untwisted sector, we already have
the vertex operator, 
which is given by the summation or the integral of \eqref{tachyon}
as explained in subsection \ref{untwist}.
However, in the twisted sectors, it is known to be difficult to 
construct vertex operators.
In this subsection, we therefore only compute the
following three point function
with the insertion of tachyon vertex operator
\begin{align}
 {}_N \langle n , \tilde n , k| : e^{i \bar p X^+ + i p X^-} :(1)
  | n , \tilde n , k \rangle_N ~,
\end{align}
where the summation or the integration over the momentum
may be needed. 
The three point function corresponds to the tachyon tadpole,
which may represent the deformation of the geometry due to the
condensation of twist field, for instance, the winding tachyon.
The higher point functions can be computed if the
extra vertex operators are in the untwisted sector.%
\footnote{See \cite{Pioline3} for the four point function of 
two twist and two untwisted fields.}

In the Hilbert space of twisted sector, we should take care of
the normal ordering of the vertex operator.
Here we adopt the normal ordering as
\begin{align}
 : e^{i \bar p X^{+} + i p X^{-}} : (z)
  = \lim_{w \to z} e^{i \bar p X^{+}} (w)
      e^{i p X^{-}} (z)
      e^{ - p \bar p [X^{u,+}_{<} (w), X^{u,-}_{>} (z) ]} ~.
\end{align}
The superscript $u$ denotes the mode expansion in the untwisted
sector, and the subscripts $>$ and $<$ represent the positive and negative
modes, respectively. In the $k$-th twisted sector, 
the vertex operator is explicitly written as
\begin{align}
 : e^{i \bar p X^{+} + i p X^{-}} :
= e^{i \bar p X^{+}_{<} + i p X^{-}_{<}} 
e^{i \bar p X^{+}_{>} + i p X^{-}_{>}}
e^{i \bar p x^{+}_0 + i p \tilde x^{-}_0} 
e^{i p x^{-}_0 + i \bar p \tilde x^{+}_0} 
 \delta_k^{ - p \bar p} ~,
\end{align}
where we have defined ($\gamma$ is the Euler number)
\begin{align}
\begin{aligned}
 \ln \delta_k &= \sum_{n=0}^{\infty} \left[ 
  -2 \frac{1}{n+1} + \frac{1}{n + \frac{k}{N}} + \frac{1}{n + 1 - \frac{k}{N}}
   \right] = 
 2 \psi (1) - \psi (\frac{k}{N}) - \psi (1-\frac{k}{N}) ~,
 \\
\psi (x) &= \frac{d}{dx} \ln \Gamma (x) 
 = - \gamma - \sum_{n = 0}^{\infty} 
   \left( \frac{1}{z+n} - \frac{1}{n+1}\right)~,
\end{aligned}
\label{delta_k}
\end{align}
and for the quasi-zero mode parts as
\begin{align}
 x^{+}_0 &= - i {\textstyle \frac{N}{k}} \alpha^+_0 z^{\frac{k}{N}} ~,
 &x^{-}_0 &= i {\textstyle \frac{N}{k}} \alpha^-_0 z^{-\frac{k}{N}} ~,
&\tilde x^{+}_0 &= i  {\textstyle \frac{N}{k}} \tilde \alpha^+_0 \bar z^{-\frac{k}{N}} ~,
&\tilde x^{-}_0 &= - i  {\textstyle \frac{N}{k}} \tilde \alpha^-_0 \bar z^{\frac{k}{N}} ~.
\end{align}
Without the excitation of quasi-zero modes, we therefore obtain
\begin{align}
 {}_N \langle k| : e^{i \bar p X^+ + i p X^-} :  (1)
  | k \rangle_N  = \delta_k^{- p \bar p} ~.
\end{align}
This factor arises due to a higher mode effect, and the effect
in Misner space was examined in \cite{Pioline3} as stringy 
fuzziness.

Now we can include the excitation of quasi-zero modes.
We need to compute
\begin{align}
 {}_N \langle k | e^{- y \alpha^-_0 - \tilde y \tilde \alpha^+_0}
e^{ \bar p \frac{N}{k} \alpha^+_0 + p \frac{N}{k} \tilde \alpha^{-}_0} 
e^{ - p \frac{N}{k} \alpha^-_0 - \bar p \frac{N}{k} \tilde \alpha^{+}_0} 
  e^{- y ' \alpha^+_0 - \tilde y ' \tilde \alpha^-_0}
  | k \rangle_N ~,
\end{align}
which can be evaluated by utilizing
Baker-Campbell-Hausdorff formula.
Gathering the quasi-zero mode and higher mode parts, we obtain
\begin{align}
 {}_N  \langle y, \tilde y , k | : e^{i \bar p X^+ + i p X^-} : (1)
  | y ' , \tilde y ' , k \rangle_N
  = e^{\frac{k}{N}(yy' + \tilde y \tilde y') 
    - \bar p (y - \tilde y') + p (y ' - \tilde y)} \delta_k^{- p \bar p} ~. 
\label{uutop}
\end{align}
If we insert $- i \gamma$ instead of $1/N$, then we obtain the
Misner space result, which is the same as the one in \cite{Pioline3}.
The Mellin transformation \eqref{Mellin} of this function is also 
possible, and the result may be expressed by Tricomi's confluent 
hyperbolic function as
\begin{align}
 {}_{\gamma}  \langle \omega^2, \tilde \omega {}^2, k | 
  : e^{i \bar p X^+ + i p X^-} : (1)
 | \omega '{}^2, {\tilde \omega}'{}^2, k \rangle_{\gamma} 
 = \xi ( - p)^{\frac{\omega^2 - \omega ' {}^2}{i k \gamma} }
 (- \bar p) ^{\frac{\tilde \omega^2 - \tilde \omega ' {}^2}{i k \gamma}}
 (i k \gamma)^{- 1 - \frac{\omega^2 + \tilde \omega^2}{i k \gamma}} 
 \nonumber \\ \times
U\left( \frac{\omega^2 }{i k \gamma} + \frac12, 
   \frac{\omega^2 - \omega '{}^2}{i k \gamma} + 1 , \frac{i p \bar p}{k \gamma} \right)
U\left( \frac{\tilde \omega^2 }{i k \gamma} + \frac12, 
   \frac{\tilde \omega^2 - \tilde \omega '{}^2}{i k \gamma} + 1 , \frac{i p \bar p}{k \gamma} \right)   
 \delta_k^{- p \bar p} ~,
\end{align}
where the integral representation
\begin{align}
 U(a,b,z) = \frac{1}{\Gamma (a)}
 \int_0^{\infty} e^{-zt} t^{a-1} (t+1)^{-a+b-1} dt
\end{align}
is used.
As discussed before, the condensation of winding tachyon may
deform the geometry near the big bang singularity in the cosmological
region. This result suggests that a stringy effect enhances the
deformed region due to the string fuzziness.

\subsection{Classical three point functions of twist fields}
\label{classical}

The wave functions corresponding to the twisted states
in ${\mathbb C}/{\mathbb Z}_N$ theory can be constructed as follows.
Because the quasi-zero modes satisfy the commutation relations
\eqref{comm_rel}, the operators may be represented as
\begin{align}
 &\alpha_0^{\pm} = - i \partial_{\mp} \pm i \frac{k}{2N} x^{\pm} ~,
 &\tilde \alpha_0^{\pm} &= - i \partial_{\mp} \mp i \frac{k}{2N} x^{\pm}  ~.
\end{align}
The definition of the vacuum state \eqref{vac_twist}
gives the condition that the corresponding wave function 
$\Psi^+_k$ should be annihilated by $\alpha^-_0$ and $\tilde \alpha^+_0$.
The condition leads that the wave function is of the Gaussian form as 
\begin{align}
 \Psi^+_k (x^{\pm} ) &= C e^{- \frac{k}{2N} x^+ x^-}  ~,
 &\Psi^-_k (x^{\pm} ) & \left( = {\Psi^+_k (x^{\pm} )}^* \right) 
 =  C^* e^{- \frac{k}{2N} x^+ x^-}  ~.
\end{align}
The normalization is fixed as $C = \sqrt{\frac{k}{N \pi}}$,
so that $\int dx^2 \Psi^-_k \Psi^+_k = 1$.
The three point function without excitation is then computed as
the overlap of the wave functions as $(k_1 = k_2 + k_3)$
\begin{align}
 C_3 = \int dx^2 \Psi^{\mp}_{k_1}\Psi^{\pm}_{k_2}\Psi^{\pm}_{k_3}
 = \frac{1}{\sqrt{\pi}} 
 \sqrt{\frac{\frac{k_2}{N} \frac{k_3}{N}}{\frac{k_1}{N}}} ~.
 \label{classical_3pt}
\end{align}
The inclusion of excitation is given by acting the creation
operators to the vacuum wave function as
\begin{align}
 \Psi^{+,n, \tilde n}_{k} = 
 ( \alpha^+_0 )^n  ( \tilde \alpha^-_0 )^{\tilde n} \Psi^+_k
  =  \textstyle (- i \partial_- + i \frac{k}{2 N} x^+ )^n  
   (- i \partial_+ +  i \frac{k}{2N} x^- )^{\bar n} \Psi^+_k ~,
\end{align}
and also $\Psi^{-,n, \tilde n}_k = ( \Psi^{+,\tilde n , n}_k )^*$.

Using the above wave functions, now we can compute the overlaps of the 
excited wave functions, for example, as
\begin{align}
 \int dx^2 \Psi^{\mp,m,0}_{k_1} \Psi^{\pm,n,0}_{k_2} \Psi^{\pm}_{k_3}
  =    \delta_{m,n}  m! \left( \frac{k_2}{N}  \right)^m 
        C_3 ~, \label{hol-hol}\\
 \int dx^2 \Psi^{\mp}_{k_1} \Psi^{\pm,n,0}_{k_2} \Psi^{\pm, 0 , \tilde n}_{k_3}
  =    \delta_{n, \tilde n}  n!
 \left( - \frac{\frac{k_2}{N} \frac{k_3}{N}}{\frac{k_1}{N}} \right)^n 
       C_3 ~. \label{hol-anti1}
\end{align}
Non-trivial overlaps may be
\begin{align}
 \int dx^2 \Psi^{\mp}_{k_1} \Psi^{\pm, n , \tilde n}_{k_2} \Psi^{\pm}_{k_3}
  =    \delta_{n, \tilde n}  n!
 \left( \frac{\frac{k_2}{N} \frac{k_3}{N}}{\frac{k_1}{N}} \right)^n 
  C_3 ~, \label{hol-anti2}
\end{align}
and 
\begin{align}
 \int dx^2 \Psi^{\mp, n \tilde n}_{k_1} \Psi^{\pm}_{k_2} \Psi^{\pm}_{k_3}
  =  0 ~,
\end{align}
where we should use
\begin{align}
\alpha^+_0 \tilde \alpha^-_0 = - \frac{k}{N}\frac{k}{N}
 x^+ x^- + \frac{k}{N} ~.
\end{align}
In the language of the basis
\begin{align}
 \Psi^{\pm , y, \tilde y}_k = \sum_{n, \tilde n} 
 \frac{(-y)^n(-\tilde y)^{\tilde n}}{n! \tilde n!} \Psi^{\pm , n , \tilde n}_{k} ~,
\end{align}
the three point functions can be summarized as
\begin{align}
 \int dx^2 \Psi^{\mp,y_1,\tilde y_1}_{k_1}  
 \Psi^{\pm,y_2,\tilde y_2}_{k_2} \Psi^{\pm,y_3,\tilde y_3}_{k_3} 
  = e^{ y_3 (\frac{k_1}{N} y_1 + \frac{k_2}{N} y_2) +
       \tilde y_3 (\frac{k_1}{N} \tilde y_1 + \frac{k_2}{N} \tilde y_2 ) 
        + \frac{\frac{k_2}{N} \frac{k_3}{N}}{\frac{k_1}{N}}
          (y_2 - y_3)(\tilde y_2 - \tilde y_3) } C_3 ~.
 \label{t3pt}
\end{align}
Replacing $1/N$ by $- i \gamma$, we reproduce the result in
\cite{Pioline3}.
In subsection \ref{ttt}, we derive the correlators in the full 
quantum level by utilizing the monodromy conditions.

\section{Correlation functions}
\label{correlators}

In the ${\mathbb C}/{\mathbb Z}_N$ orbifold model, 
the coordinate fields $X^{\pm}$ have
twisted boundary conditions \eqref{tbc}, and
the twist can be generated by the insertion of twist
field $\sigma^{\pm}_{k}$ $(k > 0)$, 
whose operator product expansions (OPEs) are
\begin{align}
\begin{aligned}
 i \partial X^+ (z) \sigma^+_{k} ( w , \bar w )
  &\sim \frac{\tau^+ (w, \bar w)}{(z-w)^{1 - \frac{k}{N}}}  ~,
 &i \partial X^- (z) \sigma^+_{k} ( w , \bar w )
  &\sim \frac{\tau^+{}' (w, \bar w)}{(z-w)^{ \frac{k}{N}}}  ~,\\
 i \bar \partial X^+ ( \bar z) \sigma^+_{k} ( w , \bar w )
  &\sim \frac{{\tilde \tau^+}{}' (w, \bar w)}{(\bar z-\bar w)^{ \frac{k}{N}}}  ~,
 &i \bar \partial X^- ( \bar z) \sigma^+_{k} ( w , \bar w )
  &\sim \frac{{\tilde \tau^+} (w, \bar w)}{(\bar z- \bar w)^{1 - \frac{k}{N}}}  ~,
  \label{OPE}
\end{aligned}
\end{align}
and similar ones for $\sigma^-_k$.
These twist fields are related to the vacuum states 
in the twisted sector by
\begin{align}
\lim_{z \to 0} \sigma_k^+ (z) | 0 \rangle_N &= |  k \rangle_N ~,
&\lim_{z \to \infty} (- z^2)^{ h_k }
 {}_N \langle 0 | \sigma_k^- (z)  
&= {}_N \langle k |
\label{vac_cor} ~,
\end{align}
where we denote the conformal weight as 
$h_k = \frac{1}{2}\frac{k}{N}(1 - \frac{k}{N})$.
We can define the twist fields with the excitation of quasi-zero modes by
repeating the OPEs \eqref{OPE} as
\begin{align}
\begin{aligned}
 : (i \partial X^+)^n ( i \bar \partial X^-)^{\tilde n} : (z,\bar z)
  \sigma_k^+ (w,\bar w) \sim 
  \frac{\sigma^{+,n,\bar n}_k (w,\bar w)}
    {(z-w)^{n(1- \frac{k}{N})} 
     (\bar z- \bar w)^{\tilde n(1- \frac{k}{N})} } ~, \\
 \sigma_k^- (z,\bar z) : (i \partial X^-)^n ( i \bar \partial X^+)^{\tilde n} : (w,\bar w)
   \sim 
  \frac{\sigma^{-,n,\bar n}_k (w,\bar w)}
    {(z-w)^{n(1- \frac{k}{N})} 
     (\bar z- \bar w)^{\tilde n(1- \frac{k}{N})} } ~.
\end{aligned}
\label{twistnn}
\end{align}
The orderings are chosen to be consistent
with the definitions \eqref{vac_cor} and \eqref{vac_twist_ket},
\eqref{vac_twist_bra}.
This definition of excited twist fields implies
$\sigma^{+,1,0}_k = \tau^+$, $\sigma^{+,0,1}_k = \tilde \tau^+$.
Note that $\tau^+{}'$ and $\tilde \tau^+ {}'$ include not 
quasi-zero modes but higher excited modes.
We can also define
\begin{align}
 \sigma^{\pm ,y,\tilde y}_k = \sum_{n, \tilde n =0}^{\infty} 
   \frac{(-y)^n (-\tilde y)^{\tilde n}}{n ! \tilde n !} 
 \sigma_k^{\pm ,n,\tilde n}~,
\label{twistxy}
\end{align}
and $\sigma^{\pm,\omega^2 , \tilde \omega^2}_k$ by replacing
$1/N \leftrightarrow -i\gamma$, $X_2 \leftrightarrow -iX_0$
and performing the Mellin transformation \eqref{Mellin}.
In the following, we will try to compute correlation functions
involving the twist fields of the form \eqref{twistxy}.

We will utilize the monodromy conditions around twist fields
as in \cite{DFMS}. This method may be applied to
any correlators, but we mainly compute four point 
functions and read off the three point functions
through the factorization. We begin with the four point
function without excitation 
\begin{align}
 Z_N(z_i , \bar z_i )
 &= \langle \prod_{i=1}^4 \sigma^{\epsilon_i}_{k_i} (z_i , \bar z_i ) 
    \rangle_N ~, 
 \label{4pt}
\end{align}
where $k_1 + k_3 = k_2 + k_4$ and
$\epsilon_1 = - \epsilon_2 = \epsilon_3 = - \epsilon_4 = - 1$.
The green functions in the presence of the twist fields are
defined as
\begin{align}
g(z,w;z_i,\bar z_i) = \frac{\langle - \partial X^+ (z) \partial X^- (w)
 \prod_{i=1}^4 \sigma^{\epsilon_i}_{k_i} (z_i, \bar z_i) \rangle_N }{
  \langle \prod_{i=1}^4 \sigma^{\epsilon_i}_{k_i} (z_i, \bar z_i) \rangle_N } ~,
\label{green_g}
\end{align}  
and
\begin{align}
h(\bar z,w;z_i , \bar z_i) = \frac{\langle - \bar \partial X^+ (\bar z) \partial X^- (w)
 \prod_{i=1}^4 \sigma^{\epsilon_i}_{k_i} (z_i, \bar z_i) \rangle_N }{
  \langle \prod_{i=1}^4 \sigma^{\epsilon_i}_{k_i} (z_i, \bar z_i) \rangle_N } ~.
\label{green_h}
\end{align} 
We also define the other green functions $\bar g (\bar z, \bar w)$
and $\bar h (z, \bar w)$ by interchanging
$\partial \leftrightarrow \bar \partial$.

The excitation of quasi-zero modes can be incorporated by
utilizing the green functions as
\begin{align}
\begin{aligned}
 &\langle \sigma_{k_1}^{-,n,0} (z_1) \sigma^{+,n,0}_{k_2} (z_2)
         \sigma^-_{k_3}(z_3) \sigma^+_{k_4}(z_4) \rangle_N \\
   &= \lim_{ z \to z_2 , w \to z_1 } 
      (z - z_2)^{n(1-\frac{k_2}{N})}
      (z_1 - w)^{n(1-\frac{k_1}{N})}
 \langle (i \partial X^+ (z))^n (i \partial X^- (w))^n
 \prod_{i=1}^4 \sigma^{\epsilon_i}_{k_i} (z_i, \bar z_i) \rangle_N \\
 &= 
 n! g ' (z_2,z_1 ; z_i , \bar z_i)^n \langle \prod_{i=1}^4 \sigma^{\epsilon_i}_{k_i} (z_i, \bar z_i) \rangle_N ~,
 \label{4pt_g^n} 
\end{aligned}
\end{align}
where we have defined
\begin{align}
 g ' (z_2,z_1; z_i , \bar z_i) = \lim_{z \to z_2, w \to z_1}
 (z - z_2)^{1-\frac{k_2}{N}}
 (z_1 - w)^{1-\frac{k_1}{N}} g (z , w ; z_i . \bar z_i) ~.
 \label{limit}
\end{align}
Recall that we have defined the excited states through the OPEs \eqref{twistnn}.
We will also use the functions $h',\bar g', \bar h'$,
which are defined in the same procedure to take the limits.
It is straightforward to convert into
the basis \eqref{twistxy} as
\begin{align}
\begin{aligned}
 &\langle \sigma_{k_1}^{-,y_1,0} (z_1) \sigma^{+,y_2,0}_{k_2} (z_2)
         \sigma^-_{k_3}(z_3) \sigma^+_{k_4}(z_4) \rangle_N
 = 
 e^{ y_2 y_1 g ' (z_2,z_1; z_i , \bar z_i)}
 \langle \prod_{i=1}^4 \sigma^{\epsilon_i}_{k_i} (z_i, \bar z_i) \rangle_N ~. 
\end{aligned}
\end{align}
Generalizing this, we can write down the four point functions
with general excitations in terms of $g',\bar g' , h' , \bar h'$ as
\begin{align}
 \langle \prod_{i=1}^4 \sigma_{k_i}^{\epsilon_i,y_i,\tilde y_i} (z_i) \rangle_N
 &= 
 e^{ y_2 y_1 g ' (z_2,z_1) + y_2 y_3 g ' (z_2,z_3) 
   + y_4 y_1 g'(z_4,z_1) + y_4 y_3 g'(z_4,z_3)}  \nonumber \\
 &\times e^{ \tilde y_1 y_1 h ' (\bar z_1,z_1) + \tilde y_1 y_3 h ' (\bar z_1,z_3) 
   + \tilde y_3 y_1 h'(\bar z_3,z_1) + \tilde y_3 y_3 h'(\bar z_3,z_3)}  \label{general4pt}\\
 &\times e^{ \tilde y_1 \tilde y_2 \bar g ' (\bar z_1,\bar z_2) 
   + \tilde y_1 \tilde y_4 \bar g ' (\bar z_1, \bar z_4) 
   + \tilde y_3 \tilde y_2 \bar g'(\bar z_3,\bar z_2) 
   + \tilde y_3 \tilde y_4 \bar g'(\bar z_3,\bar z_4)} \nonumber \\
 &\times e^{ y_2 \tilde y_2 \bar h ' (z_2,\bar z_2) + y_4 \tilde y_2 \bar h ' (z_4,\bar z_2) 
   + y_2 \tilde y_4 \bar h'(z_2,\bar z_4) + y_4 \tilde y_4 \bar h'(z_4,\bar z_4)} 
\langle \prod_{i=1}^4 \sigma^{\epsilon_i}_{k_i} (z_i, \bar z_i) \rangle_N ~. 
\nonumber
\end{align}
The explicit form will be determined through the
monodromy conditions below.

The three point functions can be read off from
the above four point function \eqref{general4pt}
though the factorization as follows.
Consider the four point function of primary fields
\begin{align}
 \langle \prod_{i=1}^4 \Phi_i (z_i , \bar z_i) \rangle ~.
\end{align}
We suppose that the primary fields $\Phi_i$ have
the conformal weights $(h_i,\bar h_i)$.
If we take the limit of $z_1 \to z_2$ and $z_3 \to z_4$,
then it is useful to use the OPEs of the primary
fields
\begin{align}
 \Phi_1 (z_1 , \bar z_1) \Phi_2 (z_2 , \bar z_2 )
  &= \sum_I \frac{C^I_{12} \Phi_I (z_2, \bar z_2)}
  {z_{12}^{\Delta_{12}} \bar z_{12}^{\bar \Delta_{12}} } ~,
 &\Phi_3 (z_3 , \bar z_3) \Phi_4 (z_4 , \bar z_4 )
  &= \sum_I \frac{C^I_{34} \Phi_I (z_4 , \bar z_4)}
  {z_{34}^{\Delta_{34}} \bar z_{34}^{\bar \Delta_{34}} } ~,
\end{align}
where $z_{12}=z_1 - z_2$, $\Delta_{12}=h_1 + h_2 - h_I$, and so on.
Therefore, in this limit, the leading term of the four point function
is given by the product of three point functions as
\begin{align}
 \langle \prod_{i=1}^4 \Phi_i (z_i , \bar z_i) \rangle 
 \sim \sum_p \frac{1}{z_{24}^{2h_p}\bar z_{24}^{2 \bar h_I}}
 \frac{C^I_{12}}{z_{12}^{\Delta_{12}} \bar z_{12}^{\bar \Delta_{12}} }
 \frac{C_{34I}}{z_{34}^{\Delta_{34}} \bar z_{34}^{\bar \Delta_{34}} } ~.
\label{factorization}
\end{align}
In the following analysis, we may set $(z_1,z_2,z_3,z_4)=(0,x,1,\infty)$,
and take the limit of $x \to 0$ or $x \to \infty$.
When we take the limit of $x \to 0$, we use the OPE between
$\Phi_2$ and $\Phi_1$, and
when we take the limit of $x \to \infty$, we use the OPE between
$\Phi_2$ and $\Phi_4$.

In the next subsection, we compute a general four point function
by following \cite{DFMS}.
After that, we deduce the three point functions
of twisted fields in subsection \ref{ttt}, which are the quantum
version of the ones in subsection \ref{classical}.
Then, we move to the three point functions with an 
untwisted fields in subsection \ref{tut}, which is found to be
consistent with the results
in subsection \ref{operator}.

\subsection{Four point functions}
\label{four}

We start from the four point function 
of twist fields without excitation \eqref{4pt},
which will be determined from
the green functions \eqref{green_g} and \eqref{green_h}
as explained below.
Utilizing the monodromy conditions \eqref{OPE}, we can fix the 
form of the green function \eqref{green_g} almost uniquely as%
\footnote{We should note that the assumption 
of integer $N$ is not needed here and below. In fact, the four point function 
\eqref{4pt} was already calculated for an irrational orbifold 
case in \cite{DAK} as a free field realization of Nappi-Witten 
model.}
\begin{align}
g(z,w; z_i , \bar z_i) = \omega_k (z) \omega_{N-k}(w)
 \left[ \frac{P(z,w,z_i)}{(z-w)^2} + A(z_i,\bar z_i)  \right]
\end{align}
with unfixed function $A(z_i,\bar z_i)$.
The classical parts $\partial X^+ (z) = \omega_k (z)$ 
and  $\partial X^- (z) = \omega_{N -k} (z)$ are determined as
\begin{align}
\begin{aligned}
 \omega_k (z) = (z-z_1)^{-\frac{k_1}{N}} 
  (z-z_2)^{- 1 + \frac{k_2}{N}} (z-z_3)^{-\frac{k_3}{N}} 
  (z-z_4)^{- 1 + \frac{k_4}{N} } ~, \\
 \omega_{N-k} (z) = (z-z_1)^{-1 + \frac{k_1}{N}} 
  (z-z_2)^{- \frac{k_2}{N}} (z-z_3)^{- 1 + \frac{k_3}{N}} 
  (z-z_4)^{- \frac{k_4}{N} } ~,
\end{aligned}
\end{align}
which reproduces the monodromy \eqref{OPE}. 
The function $P(z,w,z_i)$ is fixed to reproduce the singular
behavior  $i \partial X^+ (z) i \partial X^- (w) \sim 1/(z - w)^2$ as
\begin{align}
 &P(z,w,z_i) = 
 \left[ \textstyle \frac{k_1}{N} - \frac{k_2}{N} \right]
  (z-z_1)(z-z_2)(w-z_3)(w-z_4) \\
 & \quad + 
\textstyle \frac{k_2}{N} (z-z_1)(z-z_3)(w-z_2)(w-z_4) +
 \left[ 1 - \frac{k_4}{N} \right]
  (z-z_2)(z-z_4)(w-z_1)(w-z_3) \nonumber \\
 & \quad+
 \left[\textstyle \frac{k_3}{N} - \frac{k_2}{N} \right]
  (z-z_2)(z-z_3)(w-z_1)(w-z_4) ~. \nonumber 
\end{align}
In particular, there should be no single pole for $z \sim w$.
This function is unique up to the shift of $A(z_i,\bar z_i)$.

The function $A(z_i,\bar z_i)$ can be determined by
global monodromy conditions
\begin{align}
 \oint_{{\cal C}_i} dz \partial X^+ + 
 \oint_{{\cal C}_i} d \bar z \bar \partial X^+
 = 0 ~,
 \label{global}
\end{align}
where the contours ${\cal C}_i$ $(i=1,2)$ are taken so that
$X^{\pm}$ receive totally no phase factors around
the twist fields inside the contours.
Here we take the closed loops ${\cal C}_1$
rounding $k_2$ times around $z_1$ and $k_1$
times around $z_2$ and ${\cal C}_2$
rounding $k_3$ times around $z_2$ and $k_2$
times around $z_3$. 
The conditions \eqref{global} may be expressed as
\begin{align}
 \oint_{{\cal C}_i} dz g(z,w) + 
 \oint_{{\cal C}_i} d \bar z h(\bar z,w)
 = 0 ~,
 \label{global2}
\end{align}
where $h(\bar z,w)$ is defined in \eqref{green_h}.
This green function is also determined up to unknown function 
$B(z_i ,\bar z_i)$ as
\begin{align}
h(\bar z,w; z_i , \bar z_i) = \bar \omega_{N -k} (\bar z) \omega_{N-k}(w) B(z_i,\bar z_i) ~.
\end{align}
Because we have now two independent global monodromy conditions  
\eqref{global2},
we can uniquely fix the undetermined functions $A(z_i,\bar z_i)$ and
$B(z_i, \bar z_i) $.

Let us solve the conditions \eqref{global2}.
We first take $w \to \infty$, and then set
$z_1 = 0$, $z_2 = x$, $z_3 = 1$ and $z_4 \to \infty$.
Using the integral representation of hypergeometric function 
\eqref{int_h}, the conditions \eqref{global2} now read
\begin{align}
 &\hat A  x^{\frac{k_2}{N}-\frac{k_1}{N}} 
   \frac{\Gamma(1-\frac{k_1}{N}) \Gamma(\frac{k_2}{N})}
        {\Gamma ( 1 - \frac{k_1}{N} + \frac{k_2}{N})} F_1(x)
 + \hat B  {\bar x}^{\frac{k_1}{N}-\frac{k_2}{N}}
   \frac{\Gamma(\frac{k_1}{N})\Gamma(1- \frac{k_2}{N})}
       {\Gamma(1  - \frac{k_2}{N} + \frac{k_1}{N})} \bar G_1 ( \bar x) \nonumber \\
 & \qquad = ({\textstyle 1 - \frac{k_4}{N}}) x^{1 - \frac{k_1}{N} + \frac{k_2}{N}}
   \frac{\Gamma (1-\frac{k_1}{N}) \Gamma( 1+\frac{k_2}{N})}
        {\Gamma ( 2 - \frac{k_1}{N} + \frac{k_2}{N})} K_1(x) ~ , \\ 
 & \hat A  (1-x)^{\frac{k_2}{N}-\frac{k_3}{N}} 
   \frac{\Gamma(1-\frac{k_3}{N})\Gamma(\frac{k_2}{N})}
        {\Gamma(1-\frac{k_3}{N}+\frac{k_2}{N})} F_2( 1 - x)
 - \hat B ( 1-{\bar x})^{\frac{k_3}{N}-\frac{k_2}{N}}
   \frac{\Gamma(\frac{k_3}{N})\Gamma(1- \frac{k_2}{N})}
        {\Gamma(1-\frac{k_2}{N}+\frac{k_3}{N})} \bar G_2 ( 1 - \bar x) \nonumber \\
 & \qquad = - ({\textstyle 1 - \frac{k_4}{N}}) (1 - x)^{1 - \frac{k_3}{N} + \frac{k_2}{N}}
   \frac{\Gamma(1-\frac{k_3}{N}) \Gamma( 1+\frac{k_2}{N})}
        {\Gamma(2 - \frac{k_3}{N} + \frac{f_2}{N} )} K_2(1 - x) ~,
\end{align}
where
\begin{align}
 \textstyle
 F_1(x) = F( \frac{k_3}{N} , 1- \frac{k_1}{N} , 
             1 -\frac{k_1}{N} + \frac{k_2}{N} ; x ) ~,
 ~ F_2(1- x) = F( \frac{k_1}{N} , 1- \frac{k_3}{N} , 
             1 -\frac{k_3}{N} + \frac{k_2}{N} ; 1 - x ) ~,
 \nonumber \\  \textstyle
 \bar G_1(\bar x) = F( 1 - \frac{k_3}{N} , \frac{k_1}{N} , 
             1 + \frac{k_1}{N} - \frac{k_2}{N} ; \bar x ) ~, 
 ~ \bar G_2(1 - \bar x) = F( 1 - \frac{k_1}{N} , \frac{k_3}{N} , 
             1 -\frac{k_2}{N} + \frac{k_3}{N} ; 1 - \bar x ) ~, 
 \nonumber \\ \textstyle
 K_1(x) = F( \frac{k_3}{N} , 1- \frac{k_1}{N} , 
             2 -\frac{k_1}{N} + \frac{k_2}{N} ;  x ) ~, 
 ~ K_2(1 - x) = F( \frac{k_1}{N} , 1- \frac{k_3}{N} , 
             2 -\frac{k_3}{N} + \frac{k_2}{N} ; 1 - x ) ~. 
             \label{FGK}
\end{align}
Here we have defined 
\begin{align}
\hat A(x, \bar x) &= - \lim_{z_4 \to \infty} z_4 A(0,x,1,z_4) ~,
&\hat B(x, \bar x) &= \lim_{z_4 \to \infty} | z_4 |^{-\frac{2 k_4}{N}} 
 B(0,x,1,z_4) ~.
\end{align}
Solving these two equations we find
\begin{align}
\hat A (x, \bar x)= x (1 - x) \frac{\partial}{\partial x} \ln I (x, \bar x) 
\end{align}
with
\begin{align}
\begin{aligned}
I(x,\bar x) = |1-x|^{\frac{2}{N}(k_3 - k_2)} 
              \frac{\Gamma (1 - \frac{k_1}{N}) \Gamma (\frac{k_2}{N})
                    \Gamma ( \frac{k_3}{N}) \Gamma ( 1 - \frac{k_2}{N})}
                   {\Gamma ( 1 - \frac{k_1}{N} + \frac{k_2}{N} )
                    \Gamma (1 - \frac{k_2}{N} + \frac{k_3}{N} )}
              F_1 (x) \bar G_2 (1-\bar x) \\
            + |x|^{\frac{2}{N}(k_1 - k_2)} 
              \frac{\Gamma(1 - \frac{k_3}{N}) \Gamma( \frac{k_2}{N})
                    \Gamma( \frac{k_1}{N} )\Gamma( 1 - \frac{k_2}{N})}
                    {\Gamma(1 - \frac{k_3}{N} + \frac{k_2}{N})
                    \Gamma ( 1 - \frac{k_2}{N} + \frac{k_1}{N})}
              F_2 (1- x) \bar G_1 (\bar x) ~,
\end{aligned}
\end{align}
and
\begin{align}
 &  \hat B (x, \bar x)= x (1 - x) 
\frac{\Gamma(1 - \frac{k_1}{N})\Gamma(\frac{k_2}{N})
      \Gamma(1 - \frac{k_3}{N})\Gamma(\frac{k_2}{N})}
     {\Gamma(1-\frac{k_1}{N}+\frac{k_2}{N})
      \Gamma(1-\frac{k_3}{N}+\frac{k_2}{N})}
 I^{-1}(x, \bar x) \\
 &  \times \left\{ (1-x)^{\frac{k_2}{N}-\frac{k_3}{N}}F_2(1-x)
  \partial_x \left[ (1-x)^{\frac{k_3}{N}-\frac{k_2}{N}}F_1(x) \right]
  - x^{\frac{k_2}{N}-\frac{k_1}{N}}F_1(x)
  \partial_x \left[ x^{\frac{k_1}{N}-\frac{k_2}{N}}F_2(1-x) \right]
 \right\} ~. \nonumber
\end{align}
During the computation, we utilized the formula \ref{rel_h0}.
Inserting these functions $\hat A (x,\bar x)$ and 
$\hat B (x , \bar x)$,
we have the explicit form of the green functions
$g(z,w;x,\bar x)$ \eqref{green_g}
and $h(\bar z,w; x,\bar x)$ \eqref{green_h}.
The other two green functions 
$\bar g (\bar z , \bar w ; x , \bar x)$
and $\bar g (z , \bar w ; x , \bar x)$ are given by
replacing the arguments with bar and without bar
and $k_i \leftrightarrow N - k_i$.

{}From the expression of $\hat A(x,\bar x)$, we can read off the
four point function \eqref{4pt} as follows. A limit of
the green function \eqref{green_g} gives the correlation
function involving the energy momentum tensor as
\begin{align}
 \lim_{z \to w} \left[ g(z,w) - \frac{1}{(z-w)^2} \right] =
 \frac{\langle T(z)
 \prod_{i=1}^4 \sigma^{\epsilon_i}_{k_i} (z_i, \bar z_i) \rangle }{
  \langle \prod_{i=1}^4 \sigma^{\epsilon_i}_{k_i} (z_i, \bar z_i) \rangle } ~.
 \label{correlation_T}
\end{align}  
Since the OPE with the energy momentum tensor
reads\footnote{We denote
$h_i (= h_{k_i}) = \frac12 \frac{k_i}{N}(1 - \frac{k_i}{N})$.}
\begin{align}
 T(z) \sigma^+_{k_2} (z_2) \sim
  \frac{h_2 }{(z-z_2)^2}
  + \frac{\partial_{z_2} \sigma_{k_2} (z_2)}{z-z_2} ~,
\end{align} 
we can obtain from the correlation function \eqref{correlation_T} as
\begin{align}
 \partial_{z_2} \ln Z_N (z_i , \bar z_i) = 
 \frac{A(z_i,\bar z_i)}{(z_2-z_1)(z_2-z_3)(z_2-z_4)} -
 \frac{(1-\frac{k_1}{N})\frac{k_2}{N}}{z_2-z_1} -
 \frac{(1-\frac{k_3}{N})\frac{k_2}{N}}{z_2-z_3} -
 \frac{\frac{k_2}{N}\frac{k_4}{N}}{z_2-z_4} ~.
\end{align}
Setting $(z_1,z_2,z_3,z_4)=(0,x,1,\infty)$ and
integrating by $x$, we find
\begin{align} 
 Z_N (x,\bar x) = {\cal N} |x|^{- \frac{2 k_2}{N}(1-\frac{k_1}{N})}
       |1-x|^{- \frac{2 k_2}{N}(1-\frac{k_3}{N})}
       I^{-1}(x,\bar x)
       \label{4ptI}
\end{align}
with a constant ${\cal N}$,
which will be fixed later. Here we have defined
\begin{align}
 Z_N (x,\bar x) = \lim_{z_4 \to \infty} 
  |- z_4^2|^{h_4} Z_N (z_i , \bar z_i ) ~.
\end{align}
In this way, we have obtained the explicit form of the four point
function without excitation \eqref{4pt} as \eqref{4ptI}.

Now that we have the explicit form of the 
green functions \eqref{green_g}, \eqref{green_h} and the 
four point function \eqref{4pt}, it is easy to write
down explicitly the general four point function through
\eqref{general4pt}. 
We only need to compute the limit like \eqref{limit}.
The one in \eqref{limit} is given by
\begin{align}
\begin{aligned} 
 & g ' (z_2,z_1;z_i , \bar z_i) = e^{\pi i (-1 + \frac{k_1}{N})}
  (z_2 - z_1 )^{-\frac{k_1}{N}}  (z_2 - z_3 )^{-\frac{k_3}{N}} 
 (z_2 - z_4 )^{-1+\frac{k_4}{N}} \\ & \quad \times 
 (z_1 - z_2 )^{-\frac{k_2}{N}}  (z_1 - z_3 )^{-1+\frac{k_3}{N}} 
 (z_1 - z_4 )^{-\frac{k_4}{N}}
 \left[\textstyle A (z_i , \bar z_i)+ \frac{k_2}{N} (z_3 - z_2 )(z_2 - z_4 ) \right] 
 \\ & \quad = 
 e^{\pi i \frac{k_4}{N}} 
  x^{-\frac{k_1}{N} - \frac{k_2}{N}}
  (x-1)^{-\frac{k_3}{N}}  
  \left[ \textstyle \hat A(x,\bar x) + \frac{k_2}{N} (1 - x ) \right] ~.
\end{aligned}
\label{g'21}
\end{align}
In the last equation, we fix the positions as 
$(z_1,z_2,z_3,z_4)=(0,x,1,\infty)$.
In the same way, we obtain
\begin{align}
 g ' (z_2 , z_3 ) = - 
   x^{- \frac{k_1}{N}} (1 - x)^{- \frac{k_2}{N} - \frac{k_3}{N}}
    \left[ \hat A (x,\bar x) - \textstyle \frac{k_2}{N} x \right] ~.
\label{g'23}
\end{align}
When we take the limit of $z \to z_4$ and also $z_4 \to \infty$,
we need to define as
\begin{align}
 g ' (z_4 , z_1 ; x , \bar x) &= 
 \lim_{z_4 \to \infty} (- z_4^2)^{\frac{k_4}{N}} 
 g'(z_4 , z_1 ; 0,x,1,z_4) \label{g'41}\\
 &= -
 e^{\pi i \frac{k_4}{N}} 
   x^{- \frac{k_2}{N}}
    \left[ \hat A (x,\bar x) - \textstyle \frac{k_1}{N} 
         + \frac{k_2}{N} - \frac{k_2}{N} x \right] ~, \nonumber \\
 g ' (z_4 , z_3 ; x , \bar x) &= 
 \lim_{z_4 \to \infty} (- z_4^2)^{\frac{k_4}{N}} 
 g'(z_4 , z_3 ; 0,x,1,z_4) \label{g'43} \\ 
 &= e^{\pi i \frac{k_3}{N}} 
   (1 - x)^{- \frac{k_2}{N}}
    \left[ \hat A(x,\bar x) \textstyle 
          + \frac{k_3}{N} - \frac{k_2}{N}  x \right] ~.
          \nonumber 
\end{align}
For the green function $h(\bar z ,w)$ \eqref{green_g},
we similarly obtain
\begin{align}
\begin{aligned}
 &h ' (\bar z_1 , z_1 ) = | x |^{- \frac{2 k_2}{N}} \hat B (x,\bar x)~,
 \quad 
h ' (\bar z_1 , z_3 ) = e^{-\pi i (-1+\frac{k_3}{N})}
      [\bar x (x - 1)]^{- \frac{2 k_2}{N} } \hat B (x,\bar x)~, \\
 &h ' (\bar z_1 , z_1 ) = e^{ \pi i (-1+\frac{k_3}{N})}
      [ x (\bar x - 1)]^{- \frac{2 k_2}{N}} \hat B (x,\bar x)~, \quad
 h ' (\bar z_3 , z_3 ) = | 1- x |^{- \frac{2 k_2}{N}} \hat B(x,\bar x) ~.
\end{aligned}
\label{h'}
\end{align}
In the next two subsections, we will read off the three point functions
from the factorization, where
the limits $x \to 0$ and $x \to \infty$ of 
the above green functions are taken.

\subsection{Three point functions of excited twist fields}
\label{ttt}

The four point function can be factorized by the product of
three point functions when expanding around $x \sim 0$ or 
$x \sim \infty$ (and also $x \sim 1$) as explained before.
We begin with the four point function without excitation
\eqref{4pt}.
First we examine the factorization around $x , \bar x \sim 0$,
where we should use the OPE between the twist fields
$\sigma_{k_1}^{-}$ and $\sigma_{k_2}^{+}$.
Suppose $k_1 - k_2 = - k_3 + k_4 = k_I > 0$, where
the twist number of the intermediate field $\sigma_{k_I}^{-}$ is set by 
the twist number conservation.
Then, we can use the OPEs as
\begin{align}
\begin{aligned}
 &\sigma_{k_1}^{-} (z,\bar z) \sigma_{k_2}^{+} ( w , \bar w)
 \sim  
 \frac{C_{(-,1)(+,2)}^{(-,I)} \sigma_{k_I}^- ( w , \bar w)}
      { |z - w|^{2 h_1 + 2 h_2 - 2 h_I}}  ~, \\
 &\sigma_{k_3}^{-} ( z , \bar z) \sigma_{k_4}^{+} ( w , \bar w )
 \sim  
 \frac{C_{(-,3)(+,4)}^{(+,I)} \sigma_{k_I}^+ ( w , \bar w )}
      { | z - w |^{2 h_3 + 2 h_4 - 2 h_I}}  ~,
\end{aligned}
 \label{twistOPE}
\end{align}
which leads to the factorization of the four point function as
\begin{align}
 Z_N (x,\bar x) \sim |x|^{-2 h_1 - 2 h_2 + 2 h_I} C_{(-,1)(+,2)}^{(-,I)}
 C_{(-,3)(+,4)(-,I)} ~.
\end{align}
Here we should note that the index is raised or lowered by the 
two point function 
$\langle \sigma^-_{k} (\infty) \sigma^+_{k'}(0) \rangle = \delta_{k,k'}$.

Since the four point function is written in terms of
$I(x,\bar x)$ as in \eqref{4ptI}, 
we can read off the three point functions \eqref{twistOPE}
from the asymptotic behavior of $I(x,\bar x)$ 
around $x , \bar x \sim 0$ 
\begin{align}
\begin{aligned}
 I (x ,\bar x) &\sim
 \frac{\Gamma(1-\frac{k_1}{N})\Gamma(\frac{k_2}{N})
       \Gamma(\frac{k_3}{N})\Gamma(\frac{k_1}{N}-\frac{k_2}{N})}
      {\Gamma(\frac{k_4}{N}) \Gamma(1 + \frac{k_2}{N} - \frac{k_1}{N}) } \\
  & \qquad +  \frac{\Gamma(1-\frac{k_3}{N})\Gamma(\frac{k_1}{N})
       \Gamma(1-\frac{k_2}{N})\Gamma(\frac{k_2}{N}-\frac{k_1}{N})}
      {\Gamma(1- \frac{k_4}{N}) \Gamma(1 - \frac{k_2}{N} + \frac{k_1}{N}) }
      | x |^{\frac{2}{N}(k_1 - k_2)} ~.
\end{aligned}
\label{asymI}
\end{align}
For later convenience, we rewrite the normalization ${\cal N}$ as
\begin{align}
 {\cal N} = v \sqrt{\frac{\pi^2 \sin \frac{\pi k_4}{N}}
 {\sin \frac{\pi k_1}{N}\sin \frac{\pi k_2}{N}\sin \frac{\pi k_3}{N}}
            } ~, \end{align}
then the three point functions \eqref{twistOPE} are given as
\begin{align}
 C_{(-,1)(+,2)(+,I)} &= \sqrt{ v  \frac{ \gamma(\frac{k_1}{N})}
 {\gamma(\frac{k_2}{N}) \gamma(\frac{k_I}{N})}} ~,
 &C_{(-,3)(+,4)(-,I)} &= \sqrt{ v  \frac{ \gamma(\frac{k_4}{N})}
 {\gamma(\frac{k_3}{N}) \gamma(\frac{k_I}{N}) }} ~,
 \label{twist3}
\end{align}
where we have used the notation $\gamma(x)=\Gamma(x)/\Gamma(1-x)$.
In the large $N$ limit, we reproduce the classical expression
\eqref{classical_3pt} with the help of the relation
\begin{align}
 \frac{ \gamma(\frac{k_1}{N})}
 {\gamma(\frac{k_2}{N}) \gamma(\frac{k_I}{N}) } 
 \sim \frac{\frac{k_2}{N}\frac{k_I}{N}}{\frac{k_1}{N}} ~.
 \label{asym_3pt}
\end{align}
For $k_2 - k_1 = - k_4 + k_3 = k_I > 0$, we can similarly obtain the
factorization as
\begin{align}
 Z_N(x,\bar x) = |x|^{-2 h_1 - 2 h_2 + 2 h_I} C_{(-,1)(+,2)}^{(+,I)}
 C_{(-,3)(+,4)(+,I)} ~.
\end{align}
The three point functions are also read from the asymptotic 
behavior of $I(x,\bar x)$ as
\begin{align}
 C_{(-,1)(+,2)(-,I)} &= \sqrt{ v \frac{\gamma(\frac{k_2}{N})}
 {\gamma(\frac{k_1}{N}) \gamma(\frac{k_I}{N})}} ~,
 &C_{(-,3)(+,4)(+,I)} &= \sqrt{ v \frac{\gamma(\frac{k_3}{N})}
 {\gamma(\frac{k_I}{N}) \gamma(\frac{k_4}{N})}} ~,
\end{align}
which are the same as the previous ones \eqref{twist3}.
Since the intermediate field is in the untwisted sector for 
 $k_2 - k_1 = 0$, 
we will discuss it separately in the next subsection.

We can also obtain similar factorization for
$x,\bar x \sim \infty$,
where the OPE between $\sigma^+_{k_2}$ and $\sigma^+_{k_4}$ is used.
Since the function $I(x,\bar x)$ has the asymptotic behavior 
\begin{align}
 I(x,\bar x) \sim |x|^{- \frac{2 k_2}{N}}
 \frac{\Gamma(\frac{k_1}{N})\Gamma(\frac{k_2}{N})
       \Gamma(\frac{k_3}{N})\Gamma(1-\frac{k_1}{N}-\frac{k_3}{N})}
      {\Gamma(1-\frac{k_4}{N})\Gamma(\frac{k_1}{N}+\frac{k_3}{N})} ~,
\end{align}
the four point function \eqref{4pt} is factorized as
\begin{align}
 |- x^2|^{2 h_2} Z_N (x,\bar x)
  \sim \left| \frac{1}{x}\right|^{- 2 h_2 - 2 h_4 + 2 h_I}
  C_{(+,2)(+,4)}^{(+,I)} C_{(-,1)(-,3)(+,I)}
\end{align}
with $k_I = k_2 + k_4 = k_1 + k_3$, and the three point functions 
are read as
\begin{align}
 C_{(+,2)(+,4)(-,I)} &= \sqrt{ v \frac{\gamma(\frac{k_I}{N})}
 {\gamma(\frac{k_2}{N}) \gamma(\frac{k_4}{N})}} ~,
 &C_{(-,1)(-,3)(+,I)} &= \sqrt{ v \frac{\gamma(\frac{k_I}{N})}
 {\gamma(\frac{k_1}{N}) \gamma(\frac{k_3}{N})}} ~. 
\end{align}
These results reproduce \eqref{twist3} obtained for 
$x , \bar x \sim 0$ as well.

We move to include excitations, namely, to compute the quantum
version of three point function \eqref{t3pt}.
We start from the case of \eqref{hol-anti2},
which can be deduced from the four point function
without excitation \eqref{4pt}.
Since the asymptotic behavior of the function $I(x, \bar x)$
is given by \eqref{asymI}, non-trivial corrections are written
for $k_1 - k_2 = k_I > 0$ as
\begin{align}
 Z_N (x , \bar x ) \sim \sum_{n=0}
  | x|^{-2 h_1 - 2 h_2 + 2 h_I + \frac{2 n k_I}{N}}
  C_{(-,1)(+,2)}^{(-,I,n,n)} C_{(-,3)(+,4)(-,I,n,n)} ~,
\end{align}
where the three point functions are
\begin{align}
\begin{aligned}
 &C_{(-,1)(+,2)(+,I,n,n)}
 = n!  \left( \epsilon \frac{\gamma(\frac{k_1}{N})}
           {\gamma(\frac{k_2}{N}) \gamma(\frac{k_I}{N})} 
      \right)^{n } C_{(-,1)(+,2)(+,I)}~, \\
 &C_{(-,3)(+,4)(-,I,n,n)}
 = n! \left( \epsilon \frac{\gamma(\frac{k_4}{N})}
           {\gamma(\frac{k_3}{N}) \gamma(\frac{k_I}{N})} 
      \right)^{ n } C_{(-,3)(+,4)(-,I)} ~.
\end{aligned}
\label{epsilon_3pt}
\end{align}
The index $(+,I,n,n)$ means that the intermediate field
is an excited field $ \sigma^{+,n,n}_{k_I}$, whose normalization
is given by \eqref{twistmetric}.
The phase factor $\epsilon = \pm 1$ cannot be determined here,
so we will fix it below in another way of factorization.
This expression reproduces the classical one 
\eqref{hol-anti2} if we use  $\epsilon = + 1$ and \eqref{asym_3pt}.
We can obtain the similar result for $k_2 - k_1  > 0$.

In order to deal with other types of excitation,
we have to consider the four point functions with
quasi-zero modes.
First we consider the correlation function
\begin{align}
 Z_N = \langle \sigma^{-,n,0}_{k_1} (z_1 , \bar z_1)
         \sigma^{+,n,0}_{k_2} (z_2 , \bar z_2 )
         \sigma^{-}_{k_3} (z_3 , \bar z_3 )
         \sigma^{+}_{k_4} (z_4 , \bar z_4 ) \rangle _ N ~,
\end{align}
where we include $(i \partial X^+ (z))^n$ and 
$(i \partial X^- (w))^n$ in the four point function
\eqref{4pt} and take the limits $z \to z_2$ and $w \to z_1$.
Setting $(z_1,z_2,z_3,z_4)=(0,x,1,\infty)$ and taking the
limit $x , \bar x \sim 0$,
we can obtain the quantum counterpart of \eqref{hol-hol}
through the factorization as follows.
Since the limit of green function is given in \eqref{g'21},
the function behaves for $k_1 - k_2 = k_I > 0$ as
\begin{align}
 g ' (z_2 ,z_1) \sim
 e^{\pi i ( \frac{k_1 }{ N} - \frac{k_2 }{ N} )} 
  x^{-\frac{k_1}{N} - \frac{k_2}{N}} 
  \textstyle \frac{k_2}{N} ~.
\end{align}
Therefore, the above correlator may be factorized as
\begin{align}
 Z_N \sim x^{- \frac{n k_1}{N} - \frac{n k_2}{N}}
       |x|^{-2 h_1 - 2 h_2 + 2 h_I} C_{(-,1,n,0)(+,2,n,0)}^{(-,I)}
       C_{(-,3)(+,4)(-,I)} ~,
\end{align}
where the three point function is%
\footnote{We will neglect the phase factor like 
$e^{\pi i ( \frac{k_1}{N} - \frac{k_2}{N})}$ in the following.
This phase factor arises when we insert 
$(z_1 , z_2 , z_3 , z_4) = (0,x,1,\infty)$ in the
factorization formula \eqref{factorization} if the conformal 
weights do not match 
for holomorphic and anti-holomorphic parts.}

\begin{align}
 C_{(-,1,n,0)(+,2,n,0)(+,I)} = n! \left( \frac{k_2}{N} \right)^n 
 C_{(-,1)(+,2)(+,I)} ~.
\end{align}
This result means that the classical result \eqref{hol-hol}
does not receive any quantum corrections.
Similarly, for $k_2 - k_1 = k_I > 0$, we find
\begin{align}
 Z_N \sim x^{- \frac{n k_1}{N} - \frac{n k_2}{N}}
       | x |^{-2 h_1 - 2 h_2 + 2 h_I} C_{(-,1,n,0)(+,2,n,0)}^{(+,I)}
       C_{(-,3)(+,4)(+,I)} ~,
\end{align}
where the three point function is
\begin{align}
 C_{(-,1,n,0)(+,2,n,0)(-,I)} = n! \left( \frac{k_1}{N} \right)^n 
 C_{(-,1)(+,2)(-,I)} ~.
\end{align}
This also reproduces \eqref{hol-hol}.
We can obtain the same result even if we use the limit of 
$x , \bar x \sim \infty$. In that case, however, we cannot fix
the phase factor $\epsilon = \pm 1$ as in \eqref{epsilon_3pt}.

Next we compute the quantum three point functions of the type
\eqref{hol-anti2} to fix the phase factor in \eqref{epsilon_3pt}.
The case of \eqref{hol-anti1} follows soon.
For that purpose, it is convenient to consider the
four point function
\begin{align}
  Z_N =  \langle   \sigma_{k_1}^{-,n,n} (z_1, \bar z_1)  
   \sigma_{k_2}^+  (z_2 , \bar z_2 ) \sigma_{k_3}^-  (z_3 , \bar z_3) 
   \sigma_{k_4}^+ (z_4, \bar z_4) \rangle_N ~,
\end{align}
and take $x , \bar x \sim \infty$ with
$(z_1 , z_2 , z_3 , z_4) = (0,x,1,\infty)$.
After some computation, we obtain the limit of the green function
\eqref{h'} 
\begin{align}
  h ' (\bar z_1,z_1)
 & =  | x |^{- \frac{2 k_2}{N}} 
  \hat B (x , \bar x) ~,
  &\hat B (x , \bar x) & \sim x(x-1) |x|^{\frac{2k_2}{N}} 
  \frac{\gamma(\frac{k_I}{N})}
       {\gamma(\frac{k_3}{N})\gamma(\frac{k_1}{N})} x^{-2}
       \label{hatB}
\end{align}
with $k_I = k_1 + k_3 = k_2 + k_4 $.
Therefore, the four point function is factorized into
\begin{align}
 | - x^2 |^{2 h_2} Z_N \sim \left| \frac{1}{x} \right|^{- 2h_2 - 2h_4 + 2h_I}
      C_{(+,2)(+,4)}^{(+,I)}
      C_{(-,1,n,n)(-,3)(+,I)}
\end{align}
with the three point function
\begin{align}
 C_{(-,1,n,n)(-,3)(+,I)}
 = n! \left( \frac{\gamma(\frac{k_I}{N})}
           {\gamma(\frac{k_1}{N}) \gamma(\frac{k_3}{N})} 
      \right)^{n} C_{(-,1)(-,3)(+,I)} ~.
\end{align}
Notice that we can fix the phase factor in \eqref{epsilon_3pt}
as $\epsilon = 1$.

The other three point function of the type \eqref{hol-anti1}
can be obtained in a similar way. Here we consider
\begin{align}
  Z_N =  \langle   \sigma_{k_1}^{-,n,0} (z_1, \bar z_1)  
   \sigma_{k_2}^+  (z_2 , \bar z_2 ) \sigma_{k_3}^{-,0,n}  (z_3 , \bar z_3) 
   \sigma_{k_4}^+ (z_4, \bar z_4) \rangle ~,
\end{align}
and take $x , \bar x \sim \infty$ with
$(z_1 , z_2 , z_3 , z_4) = (0,x,1,\infty)$.
Making use of the behavior of the green function involved \eqref{h'}
\begin{align}
  h ' (\bar z_1,z_3)
 &\sim - | x |^{- \frac{2 k_2}{N}} 
  \hat B (x , \bar x) ~,
    &\hat B (x , \bar x) & \sim |x|^{\frac{2k_2}{N}} 
  \frac{\gamma(\frac{k_I}{N})}
       {\gamma(\frac{k_3}{N})\gamma(\frac{k_1}{N})} ~,
\end{align}
the four point function can be 
factorized as
\begin{align}
 | - x^2 |^{2 h_2} Z_N \sim \left| \frac{1}{x} \right|^{- 2h_2 - 2h_4 + 2h_I}
      C_{(+,2)(+,4)}^{(+,I)}
      C_{(-,1,n,0)(-,3,0,n)(+,I)}
\end{align}
with the three point function
\begin{align}
 C_{(-,1,n,0)(-,3,0,n)(+,I)}
 = n! \left( - \frac{\gamma(\frac{k_I}{N})}
           {\gamma(\frac{k_1}{N}) \gamma(\frac{k_3}{N})} 
      \right)^{n} C_{(-,1)(-,3)(+,I)} ~.
\end{align}
This reproduces the classical result \eqref{hol-anti1} 
if we apply \eqref{asym_3pt}.

Now we have every non-trivial three point functions
with the excitation of quasi-zero mode.
In the basis of \eqref{xyk}, we can express for $k_1 = k_2 + k_3$ as
\begin{align}
 &{}_N \langle y_1 , \tilde y_1 , \mp k_1 | \sigma^{\pm,y_2 ,\tilde y_2}_{k_2}
  | y_3 , \tilde y_3 , \pm k_3 \rangle_N \nonumber \\
   &\qquad = 
   e^{ y_3 (\frac{k_1}{N} y_1 + \frac{k_2}{N} y_2) +
        \tilde y_3 (\frac{k_1}{N} \tilde y_1 + \frac{k_2}{N} \tilde y_2 ) 
        + \frac{\gamma \left(\frac{k_1}{N}\right)}
        {\gamma \left(\frac{k_2}{N}\right) \gamma \left(\frac{k_3}{N}\right)}
          (y_2 - y_3)(\tilde y_2 - \tilde y_3) } C_{(-,1)(+,2)(+,3)} ~ .
\label{ttt_3pt}
\end{align}
This expression coincides with the one in \cite{Pioline3},
where they applied
an analytic continuation to the result of Nappi-Witten model 
\cite{DAK,CFS,BDKZ}.
Our method does not only reproduce their results, but gives
a way to compute more generic correlation functions.
For instance, we have obtained a generic four point function
in \eqref{general4pt}.
The other correlation functions can be computed
by following our method.

\subsection{Three point functions with an untwisted field}
\label{tut}

In this section, we rederive the three point functions
with two twisted states and one untwisted state.
The correlation functions were already obtained in
subsection \ref{operator} in operator formalism,
but we will show that our method correctly
reproduces the results.
We consider the four point function \eqref{4pt} without
excitation for $k_1 = k_2 = k$ and $k_3 = k_4 = l$
and take the limit of $x,\bar x \sim 0$.
Here we should note that the asymptotic behavior of the 
hypergeometric functions becomes
\begin{align}
\begin{aligned}
 F_2 (1-x) \sim \frac{\Gamma (1 -  \frac{l}{N} + \frac{k}{N})}
                    { \Gamma (\frac{k}{N}) \Gamma(1 - \frac{l}{N})}
          \left[ 2 \psi (1) - \textstyle \psi (\frac{k}{N}) 
                 - \psi (1- \frac{l}{N}) - \ln x \right] ~,\\
 \bar G_2 (1- \bar x) \sim \frac{\Gamma (1 - \frac{k}{N} + \frac{l}{N})}
                    { \Gamma (\frac{l}{N}) \Gamma(1 - \frac{k}{N})}
          \left[ 2 \psi (1) - \textstyle \psi (\frac{l}{N}) 
                 - \psi (1- \frac{k}{N}) - \ln \bar x \right]  ~,
\end{aligned}
\end{align}
and hence the function $I(x,\bar x)$ behaves as
\begin{align}
 I ( x ,\bar x) \sim   \frac{\pi}{\sin  \frac{ \pi k}{N}}
   \ln \frac{\delta_k \delta_l}{|x|^2} ~,
   \label{beI}
\end{align}
where we have used the notation of $\delta_n$ in \eqref{delta_k}.
Using a formula
\begin{align}
\left(- 2 \ln \frac{|x|}{\sqrt{\delta_k \delta_l}}\right)^{-n-1}
= \frac{1}{2 \pi n!} \int d p d \bar p (p \bar p)^n 
  \left( \frac{|x|}{\sqrt{\delta_k \delta_l}}\right)^{ 2 p \bar p} ~,
  \label{log2int}
\end{align}
we find
\begin{align}
 Z_N (x,\bar x)= v \frac{|x|^{- 4 h_k}}
              {-2\ln\frac{|x|}{\sqrt{\delta_k \delta_l}}}
     = |x|^{- 4 h_k}
       \int \frac{ d p d \bar p }{ (2 \pi )^2 } 
   \left( \frac{|x|}{\sqrt{\delta_k \delta_l}}\right)^{ 2 p \bar p} ~.
   \label{ttu}
\end{align}
The above equation \eqref{ttu} implies that the three point functions
are
\begin{align}
 C_{(-,k)(+,k)p}&=\delta_k^{- p \bar p } ~,
 &C_{(-,l)(+,l)p}&=\delta_l^{- p \bar p } ~.
\end{align}
The normalization is set as%
\footnote{This does not contradict the classical result
\eqref{classical_3pt} where $v=1/\pi$ in this notation.
This can be seen from the fact that it takes 
$v=\frac{1}{\alpha ' \pi}$ if $\alpha ' =2$ is written explicitly.}
$v=\frac{1}{2\pi}$ such that $C_{(-,k)(+,k)0}=1$.
Here we should recall that in the untwisted sector 
we can map the basis of the flat space into the orbifold 
one as in subsection \ref{untwist}.

We then include the excitation of quasi-zero modes. 
We first compute the four point function
\begin{align}
Z_N = \langle \sigma^-_{k}(z_1 , \bar z_1) 
              \sigma^{+,m,0}_{k}(z_2 , \bar z_2)
          \sigma^{-,m,0}_{l}(z_3 , \bar z_3)
            \sigma^+_{l}(z_4 , \bar z_4) \rangle_N ~,
\end{align}
and take the limit of $x , \bar x \sim 0$ with 
$(z_1 , z_2 , z_3 , z_4) = (0,x,1,\infty)$.
We utilize the green function \eqref{g'23}, which behaves as
\begin{align}
 g ' (z_2, z_3) & =  - (-1)^{\frac{l}{N}}
    x^{-\frac{k}{N}} (1-x)^{-\frac{k}{N}-\frac{l}{N}} 
    [ \hat A (x, \bar x) - { \textstyle \frac{k}{N}} x ] ~,
    &A (x, \bar x) &\sim \frac{1}{- \ln \frac{\delta_k \delta_l}{|x|^2}} ~.
\end{align}
With the help of this behavior and the formula \eqref{log2int},
we can obtain the factorization as
\begin{align}
 Z_N &\sim \frac{v x^{-\frac{m k}{N}} |x|^{ - 4 h_k}}
            {- 2 \ln \frac{|x|}{\sqrt{\delta_k \delta_l}}}
  m !  \left(\frac{1}{ - 2 \ln \frac{|x|}{\sqrt{\delta_k \delta_l}}}\right)^{m}  
 \nonumber \\
  &=
 x^{-\frac{m k}{N}} |x|^{ - 4 h_k}
  \int \frac{d p d \bar p}{(2 \pi)^2} 
  C_{(-,k)(+,k,m,0)}^{p} C_{(-,l,m,0)(+,l)p}
  ~,
\end{align}
where the three point functions are%
\footnote{The phase factor cannot be fixed only from the
factorization. We use the one consistent with the result
in operator formalism.}
\begin{align}
 C_{(-,k)(+,k,m,0)p} &= 
 ( - p )^m 
 \delta^{- p \bar p}_k ~,
 &C_{(-,l,m,0)(+,l)p} &= 
  ( - {\bar p })^m 
 \delta^{- p \bar p }_l ~.
\end{align}

We next compute a more generic four point function
\begin{align}
Z_N = \langle \sigma^{-,n,0}_{k}(z_1,\bar z_1) 
          \sigma^{+,n+m,0}_{k}(z_2,\bar z_2) 
           \sigma^{-,m,0}_{l}(z_3 , \bar z_3) 
            \sigma^+_{l}(z_4 , \bar z_4) \rangle _N ~,
\end{align}
and take the limit of $x , \bar x \sim 0$ with 
$(z_1 , z_2 , z_3 , z_4) = (0,x,1,\infty)$.
Here we use the limit of the green function \eqref{g'21} as well.
{}From the similar computation, we obtain
\begin{align}
 Z_N &\sim 
 \frac{ v x^{-\frac{(2 n + m) k }{N}} |x|^{ - 4 h_k}}
    {- 2 \ln \frac{|x|}{\sqrt{\delta_k \delta_l}}}
   (n + m) ! \left( \frac{k}{N} 
  - \frac{1}{- 2 \ln \frac{|x|}{\sqrt{\delta_k \delta_l}}} \right)^n
  \left(
  \frac{1}{- 2 \ln \frac{|x|}{\sqrt{\delta_k \delta_l}}} \right)^{m} \\
    &=
 x^{-\frac{(2 n + m) k }{N}} |x|^{ - 4 h_k } 
   \int \frac{d p d \bar p}{(2 \pi)^2} {|x|}^{2 p \bar p}
  C_{(-,k,n,0)(+,k,n+m,0)}^{p}  
  C_{(-,l,m,0)(+,k)p}
\end{align}
with the three point function
\begin{align}
 C_{(-,k,n,0)(+,k,n+m,0)p} = 
 (n+m)! \left( \frac{k}{N} \right)^{n+m}
 { \bar p }^{ - m }
 L_{n + m}^{(-m)} \left( \frac{N p \bar p}{k} \right) 
 \delta^{- p \bar p }_k ~.
\end{align}
The definition of generalized Laguerre polynomial
\begin{align}
 L_n^{(m)} (x) = \sum_{l=0}^{n} (-1)^l 
 \left( \begin{matrix} n+m\\n-l\end{matrix}\right)
 \frac{x^l}{l!} 
\end{align}
was used in the above coefficient.

In the similar manner, we can show that 
the three point functions of the type like
$C_{(-,k,n,\bar n)(+,k)p}$ vanishes from the behavior of
the green function $h'(\bar z_1, z_1)$. 
Combined with the anti-holomorphic part, we therefore conclude that
\begin{align}
 {}_N \langle y_1 ,\tilde y_1 , k | : e^{i \bar p X^+ + i p X^-} :
      | y_2 , \tilde y_2 , k \rangle_N = 
  e^{\frac{k}{N}(y_1 y_2 + \tilde y_1 \tilde y_2 ) 
 - \bar p (y_1 - \tilde y_2) + p (y_2 - \tilde y_1)  } ~,
\label{tut_3pt}
\end{align}
which reproduces the result in operator formalism 
\eqref{uutop}. The generating function for the generalized
Laguerre polynomial
\begin{align}
 e^{xy + r (x - y)}
 = \sum_{n,l=0}^{\infty} \frac{{r}^{n-l}}{l !} L_l^{(n-l)} (r^2) x^n y^l 
\end{align}
was used to summarize in the above form.

\section{Conclusion and discussions} 
\label{conclusion}

In this paper, we have developed a general method to
compute correlation functions in Misner space.
Even though the correlators with less than two twist fields
can be computed in operator formalism as
in subsection \ref{operator},
we need a general technique to compute 
those with more twist fields.
In order to perform the path integral to compute correlation
functions, we may have to deal with an Euclidean target space.
We therefore perform the Wick rotation to the Lorentzian orbifold,
and relate to the  ${\mathbb C}/{\mathbb Z}_N$ orbifold theory.
It is a subtle problem to determine the spectrum in Misner space,
and we utilized the Mellin transformation \eqref{Mellin}
to construct a map from the spectrum in the  
${\mathbb C}/{\mathbb Z}_N$ orbifold.
The general method to compute correlators in the 
${\mathbb C}/{\mathbb Z}_N$ orbifold was already
developed in \cite{DFMS,HV} by making use of the
monodromy conditions.
We have computed a general four point function 
\eqref{general4pt}, where
the excitation of quasi-zero modes are included
by utilizing the green functions in the presence of four
twist fields. 
The three point functions have been read off
from the four point function through the factorization,
and we obtained \eqref{ttt_3pt} for those of twist fields and
\eqref{tut_3pt} for those with an untwisted field.

A motivation to compute the correlation functions is to
investigate the role of winding strings in Misner space,
in particular, the relation to the (possible) resolution of
the big crunch/big bang singularity.
Let us first see the condensation of tachyonic
strings wrapped on the whole whisker regions \eqref{winding_tachyon}.
{}From the viewpoint of worldsheet theory, 
the condensation is described by the deformation of twist fields
\begin{align}
 S = S_0 + \alpha_k \int d^2 z V_{k} + \alpha_{-k} \int d^2 z V_{- k} 
 ~,
\end{align}
where $V_{\pm k}$ denote the twist fields corresponding
to the winding tachyon fields.
The change of geometry may be read from the graviton scattering
\begin{align}
 \langle G_{\mu \nu} G^{\mu \nu} \rangle_{\alpha}
  =  \langle G_{\mu \nu} G^{\mu \nu} \rangle_0
   + | \alpha_k |^2 
    \langle V_{-k} G_{\mu \nu} G^{\mu \nu} V_k \rangle_0 
   + \cdots ~,
   \label{change}
\end{align}
where the correlators are given by the overlaps of the
corresponding wave functions with the stringy 
correction like in \eqref{uutop}.
Therefore, in the region where the tachyon 
fields are localized, the graviton modes are frozen out due
to the large mass term or the deformation of
the background geometry. 
Since the tachyon fields are wrapped over the whisker regions,
the regions with closed time-like curves may be excised
as suggested in \cite{Hagedorn}.
Moreover, the singularity in the cosmological regions
may be resolved by the the same effects \cite{MS}
possibly with the enhancement by string effects.

The effective action for the winding strings may be
constructed from the correlation functions of twist fields.
We have computed a general four point function of the form
\begin{align}
 \langle V_{-k_1} V_{k_2} V_{-k_3} V_{k_4} \rangle ~,
\end{align}
which is related to the quartic term of the winding string
interaction.
Our result \eqref{general4pt} implies that this
quantity is finite, which may lead the
following implications.
Consider the $2 \to 2$ scattering of winding strings.
The finiteness of the scattering means that the winding
string does not feel the singular property of the background
contrary to strings in the untwisted sector.
Moreover, we may be able to apply to the tachyon condensation.
In Euclidean orbifold case, it was shown in \cite{APS} that the condensation 
of localized tachyon changes the geometry, say, from 
${\mathbb C}/{\mathbb Z}_N$ into ${\mathbb C}/{\mathbb Z}_M$
with $M < N$.
The height of tachyon potential is proposed
to be the same as the difference of the geometry volume
\cite{Dabholkar}, and this proposal was investigated 
by means of string field theory \cite{OZ,BR} and
effective field theory \cite{DIR,HR}.
Since we found the quartic term is finite, it has a meaning to 
study whether a similar situation would happen or not.

We would like to investigate the following problems.
An interesting problem is whether the tachyon condensation
would separate the big crunch region and the big bang region.
In order to answer this question, we may need to develop a way to sum 
over all contribution in the abbreviation of \eqref{change}. 
We could expect that this investigation gives insights to the 
property of the tachyon state \cite{MS,NRS}.
It is also worth applying our method to more general
backgrounds. In particular, the parabolic or null orbifold 
case \cite{LMS1,LMS2} seems interesting because of the
presence of supersymmetry. We hope that we could report 
on this subject in near future.
The holographic dual description is also important because
it might give a non-perturbative picture of the resolution of the
singularity as mentioned before.

\subsection*{Acknowledgement}

We would like to thank N.~Iizuka, I.~Papadimitriou, S.~Ribault, 
V.~Schomerus and J.~Teschner for useful discussions.
This work is supported by JSPS Postdoctoral Fellowships for 
Research Abroad H18-143.


\appendix

\section{Formula for hypergeometric function}

The hypergeometric function is defined as
\begin{align}
 F(\alpha,\beta,\gamma ; z ) &= \sum_{n=0}^{\infty}
  \frac{(\alpha)_n (\beta)_n}{(\gamma)_n} \frac{z^n}{n!} ~,
  &(\alpha)_n &= \frac{\Gamma(\alpha +n)}{\Gamma(\alpha)} ~,
\end{align}
and the integral expression is given by
\begin{align}
 F(\alpha,\beta,\gamma;z) = 
 \frac{\Gamma(\gamma)}{\Gamma(\beta)\Gamma(\gamma - \beta)}
 \int_0^1 dt t^{\beta -1} (1-t)^{\gamma - \beta - 1} (1-tz)^{-\alpha} ~.
 \label{int_h}
\end{align}
We use the following relation involving the derivative
\begin{align}
 &\gamma(1-z) \partial_z  F( \alpha,\beta,\gamma;z) \nonumber \\ 
 &\qquad = (\gamma - \alpha)(\gamma - \beta) F(\alpha,\beta,\gamma+1;z)
 + \gamma(\alpha+\beta-\gamma)F(\alpha,\beta,\gamma;z) ~.
\label{rel_h0}
\end{align}

In order to see the asymptotic behavior of hypergeometric function,
the following relations are useful as
\begin{align}
 F(\alpha,\beta,\gamma;z)&=
 \frac{\Gamma(\gamma)\Gamma(\alpha + \beta - \gamma)}
      {\Gamma(\alpha)\Gamma(\beta)}
      (1-z)^{\gamma - \alpha - \beta}
       F(\gamma - \alpha,\gamma-\beta,\gamma-\alpha-\beta+1;1-z)
\nonumber\\
&+\frac{\Gamma(\gamma)\Gamma(\gamma - \alpha - \beta)}
      {\Gamma(\gamma - \alpha)\Gamma(\gamma - \beta)}
       F(\alpha,\beta,\alpha+\beta-\gamma +1;1-z) ~,
\label{rel_h1}
\end{align}
\begin{align}
 F(\alpha,\beta,\gamma;z)&=
 \frac{\Gamma(\gamma)\Gamma(\beta - \alpha)}
      {\Gamma(\beta)\Gamma(\gamma - \alpha)}
      (-z)^{- \alpha}
       F(\alpha,\alpha-\gamma+1,\alpha-\beta+1;\frac{1}{z})
\nonumber\\
&+\frac{\Gamma(\gamma)\Gamma(\alpha - \beta)}
      {\Gamma(\alpha)\Gamma(\gamma - \beta)}
      (-z)^{- \beta}
       F(\beta,\beta-\gamma+1,\beta-\alpha+1;\frac{1}{z}) ~,
\label{rel_h2} \\
 F(\alpha,\beta,\gamma;z)&=
 \frac{\Gamma(\gamma)\Gamma(\beta - \alpha)}
      {\Gamma(\beta)\Gamma(\gamma - \alpha)}
      (1-z)^{- \alpha}
       F(\alpha,\gamma-\beta,\alpha-\beta+1;\frac{1}{1-z})
\nonumber\\
&+\frac{\Gamma(\gamma)\Gamma(\alpha - \beta)}
      {\Gamma(\alpha)\Gamma(\gamma - \beta)}
      (1-z)^{- \beta}
       F(\beta,\gamma-\alpha,\beta-\alpha+1;\frac{1}{1-z}) ~.
\label{rel_h3}
\end{align}
Let us apply these relations to the functions defined in \eqref{FGK}.
For $x, \bar x \sim 0$, it is convenient to rewrite as
\begin{align}
 F_2(1-x) &=
 \frac{\Gamma(1+\frac{k_2}{N}-\frac{k_3}{N})
       \Gamma(\frac{k_1}{N}-\frac{k_2}{N})}
      {\Gamma(\frac{k_1}{N})\Gamma(1-\frac{k_3}{N})}
x^{\frac{k_2}{N}-\frac{k_1}{N}}
 F( \textstyle 1-\frac{k_4}{N},\frac{k_2}{N},1+\frac{k_2}{N}-\frac{k_1}{N};x)  \nonumber \\
 &+ \frac{\Gamma(1+\frac{k_2}{N}-\frac{k_3}{N})
       \Gamma(\frac{k_2}{N}-\frac{k_1}{N})}
      {\Gamma(1-\frac{k_4}{N})\Gamma(\frac{k_2}{N})}
 F( \textstyle \frac{k_1}{N},1-\frac{k_3}{N},1+\frac{k_1}{N}-\frac{k_2}{N};x) ~,\\
 \bar G_2(1- \bar x) &=
 \frac{\Gamma(1+\frac{k_3}{N}-\frac{k_2}{N})
       \Gamma(\frac{k_2}{N}-\frac{k_1}{N})}
      {\Gamma(\frac{k_3}{N})\Gamma(1-\frac{k_1}{N})}
{\bar x}^{\frac{k_1}{N}-\frac{k_2}{N}} 
 F( \textstyle \frac{k_4}{N},1-\frac{k_2}{N},1+\frac{k_1}{N}-\frac{k_2}{N};\bar x) \nonumber \\
 &+ \frac{\Gamma(1+\frac{k_3}{N}-\frac{k_2}{N})
       \Gamma(\frac{k_1}{N}-\frac{k_2}{N})}
      {\Gamma(\frac{k_4}{N})\Gamma(1-\frac{k_2}{N})}
 F( \textstyle 1-\frac{k_1}{N},\frac{k_3}{N},1+\frac{k_2}{N}-\frac{k_1}{N};\bar x) ~,
\end{align}
and for $x, \bar x \sim \infty$
\begin{align}
 F_1(x) &=
 \frac{\Gamma(1+\frac{k_2}{N}-\frac{k_1}{N})
       \Gamma(1-\frac{k_1}{N}-\frac{k_3}{N})}
      {\Gamma(1-\frac{k_1}{N})\Gamma(1-\frac{k_4}{N})}
(-x)^{-\frac{k_3}{N}}
 F( \textstyle \frac{k_3}{N},\frac{k_4}{N},\frac{k_1}{N}+\frac{k_3}{N};\frac{1}{x})  
\\
 &+ \frac{\Gamma(1+\frac{k_2}{N}-\frac{k_1}{N})
       \Gamma(-1+\frac{k_1}{N}+\frac{k_3}{N})}
      {\Gamma(\frac{k_3}{N})\Gamma(\frac{k_2}{N})}
(-x)^{-1+\frac{k_1}{N}}
 F( \textstyle 1-\frac{k_1}{N},1-\frac{k_2}{N},2-\frac{k_1}{N}-\frac{k_3}{N};\frac{1}{x}) ~,
 \nonumber
\end{align}
\begin{align}
 \bar G_1(\bar x) &=
 \frac{\Gamma(1+\frac{k_1}{N}-\frac{k_2}{N})
       \Gamma(-1+\frac{k_1}{N}+\frac{k_3}{N})}
      {\Gamma(\frac{k_1}{N})\Gamma(\frac{k_4}{N})}
{(- \bar x)}^{-1 + \frac{k_3}{N}} 
 F( \textstyle 1-\frac{k_3}{N},1-\frac{k_4}{N},2-\frac{k_1}{N}-\frac{k_2}{N};\frac{1}{\bar x}) 
 \nonumber \\
 &+ \frac{\Gamma(1+\frac{k_1}{N}-\frac{k_2}{N})
       \Gamma(1-\frac{k_1}{N}-\frac{k_3}{N})}
      {\Gamma(1-\frac{k_3}{N})\Gamma(1-\frac{k_2}{N})}
{(- \bar x)}^{- \frac{k_1}{N}} 
 F( \textstyle \frac{k_1}{N},\frac{k_2}{N},\frac{k_1}{N}+\frac{k_3}{N};\frac{1}{\bar x}) ~,
\end{align}
and
\begin{align}
 F_2(1-x) &=
 \frac{\Gamma(1+\frac{k_2}{N}-\frac{k_3}{N})
       \Gamma(1-\frac{k_1}{N}-\frac{k_3}{N})}
      {\Gamma(1-\frac{k_3}{N})\Gamma(1-\frac{k_4}{N})}
x^{-\frac{k_1}{N}}
 F( \textstyle \frac{k_1}{N},\frac{k_2}{N},\frac{k_1}{N}+\frac{k_3}{N};\frac{1}{x})  
\\
 &+ \frac{\Gamma(1+\frac{k_2}{N}-\frac{k_3}{N})
       \Gamma(-1+\frac{k_1}{N}+\frac{k_3}{N})}
      {\Gamma(\frac{k_1}{N})\Gamma(\frac{k_2}{N})}
 x^{-1+\frac{k_3}{N}}
 F( \textstyle 1-\frac{k_3}{N},1-\frac{k_4}{N},2-\frac{k_1}{N}-\frac{k_3}{N};\frac{1}{x}) ~,
 \nonumber
\end{align}
\begin{align}
 \bar G_2(1 - \bar x) &=
 \frac{\Gamma(1+\frac{k_3}{N}-\frac{k_2}{N})
       \Gamma(-1+\frac{k_1}{N}+\frac{k_3}{N})}
      {\Gamma(\frac{k_3}{N})\Gamma(\frac{k_4}{N})}
{\bar x}^{-1 + \frac{k_1}{N}} 
 F( \textstyle 1-\frac{k_1}{N},1-\frac{k_2}{N},2-\frac{k_1}{N}-\frac{k_3}{N};\frac{1}{\bar x}) 
 \nonumber \\
 &+ \frac{\Gamma(1+\frac{k_3}{N}-\frac{k_2}{N})
       \Gamma(1-\frac{k_1}{N}-\frac{k_3}{N})}
      {\Gamma(1-\frac{k_1}{N})\Gamma(1-\frac{k_2}{N})}
{ \bar x}^{- \frac{k_3}{N}} 
 F( \textstyle \frac{k_3}{N},\frac{k_4}{N},\frac{k_1}{N}+\frac{k_3}{N};\frac{1}{\bar x}) ~.
\end{align}

\baselineskip=11pt

\providecommand{\href}[2]{#2}\begingroup\raggedright\endgroup

\end{document}